\documentstyle[12pt]{article}

\newcommand{\mysection}{\setcounter{equation}{0}\section}

\begin{document}
\vskip 0.2cm
\hfill{NIKHEF/96-018} 
\vskip 0.2cm
\hfill{ITP-SB-96-16} 
\vskip 0.2cm
\hfill{INLO-PUB-08/96} 
\vskip 0.2cm
\centerline{\large\bf {$O(\alpha_s^2)$ corrections to polarized  }}
\centerline{\large\bf {heavy flavour production at $Q^2\gg m^2$ }}
\vskip 0.2cm
\centerline {\sc M. Buza \footnote{supported by the Foundation 
for Fundamental Research on Matter (FOM)}}
\centerline{\it NIKHEF/UVA,}
\centerline{\it POB 41882, NL-1009 DB Amsterdam,}
\centerline{\it The Netherlands.}
\vskip 0.2cm 
\centerline {\sc Y. Matiounine and J. Smith \footnote{partially supported 
under the contract NSF 93-09888}}
\centerline{\it Institute for Theoretical Physics,}
\centerline{\it State University of New York at Stony Brook,}
\centerline{\it New York 11794-3840, USA.}
\vskip 0.2cm
\centerline {\sc W.L. van Neerven}
\centerline{\it Instituut-Lorentz,}
\centerline{\it University of Leiden,}
\centerline{\it PO Box 9506, 2300 RA Leiden,}
\centerline{\it The Netherlands.}
\vskip 0.2cm
\centerline{August 1996}
\vskip 0.2cm
\centerline{\bf Abstract}
\vskip 0.3cm

In this paper we present the analytic form of the heavy flavour coefficient 
functions for polarized deep inelastic lepton-hadron scattering.
The expressions are valid in the kinematical regime $Q^2\gg m^2$ where
$Q^2$ and $m^2$ stand for the masses squared of the 
virtual photon and heavy quark
respectively. Using these coefficient functions we have computed the
next-to-leading order $\alpha_s$ corrections to polarized charm production
at HERA collider energies, where both the electron and proton beams are 
polarized.
We also give an estimate of these corrections at fixed target experiments
where the typical $Q^2$ values are much smaller than at HERA.

\vfill
\newpage

\mysection{Introduction}
\newcommand{\be}{\begin{eqnarray}}
\newcommand{\ee}{\end{eqnarray}}

Apart from another test of perturbative QCD, deep inelastic
electroproduction of charm 
quarks leads to important information about the gluon density 
inside the proton \cite{ali1}, \cite{vogt}, \cite{H1}, \cite{ZEUS}. 
This is because the dominant production mechanism
is represented by the (virtual) photon-gluon fusion 
process \cite{witten} which
is the only one appearing in the Born approximation.
Although in higher order of the strong coupling constant $\alpha_s$
other subprocesses will also contribute it turns out that the above picture
remains essentially unaltered.

Until now almost all attention was paid to unpolarized charm production.
However in future fixed target \cite{gkm} as well as collider \cite{vwh},
\cite{jb} experiments one
is also interested in charm production in polarized deep inelastic
lepton-hadron scattering (the case of photoproduction has recently
been discussed in \cite{fr} and \cite{sv}).
Like in the unpolarized case one is particularly interested in the gluon 
density since it plays an important role in the description of 
the longitudinal spin structure function
$g_1(x,Q^2)$. Here $x$ denotes the Bjorken scaling variable and $Q^2$ is
the mass squared of the virtual photon exchanged between the
lepton and the hadron. Contrary to unpolarized charm 
electroproduction, where the cross-section is already calculated up to 
next-to-leading order (NLO), only the Born approximation exists for 
the polarized case \cite{daw}, \cite{grv1}, \cite{v1}.
The calculation of the NLO corrections will be as difficult as that
for unpolarized charm electroproduction in \cite{lrsn1}, which could be only
done in a semi-analytic way. This is because the cross sections for the
parton subprocesses involve four dimensional phase space integrals where
the integrations over the azimuthal and polar angles can be performed
analytically. The two remaining integrations have to be done numerically.
After mass factorization the resulting coefficient functions are 
folded with parton densities so that one finally has to do 
integrations over three variables. Furthermore the LO parton cross sections
for $F_L$ and $F_2$ in the unpolarized case are positive definite
in contrast to those for the spin structure function $g_1$. After mass
factorization the positive definiteness does not apply to the NLO coefficient
functions in $F_L$ and $F_2$ anymore and additional positive and negative
parts appear for those in $g_1$. This leads to large
cancellations in the numerical integrations which will particularly 
complicate the computation of $g_1$. 
Hence the semi-analytic calculation of $g_1$
will be harder than that already carried out for $F_L$ and $F_2$ in 
\cite{lrsn1}. Therefore it is important to have an analytic form of the heavy
quark coefficient functions in some kinematical regime that can serve as a 
check on the exact $O(\alpha_s^2)$ calculations which have still to be done 
for the spin structure function $g_1$. Fortunately, as has been
shown for $F_L$ and $F_2$ in \cite{bmsmn}, one can obtain analytic 
expressions for the heavy quark coefficient functions in the asymptotic region
$Q^2\gg m^2$.
Here the asymptotic formulae for the heavy quark coefficient functions
could be inferred from the operator matrix elements (OME's) and the
light parton coefficient functions so that one did not have to resort to
cumbersome calculations of loop- and phase-space integrals.\\
As far as phenomenological applications are concerned, 
the asymptotic heavy quark coefficient functions will be
a good approximation at HERA collider energies, where both the
electron and the proton beams are polarized, because there will be
events at $Q^2 \gg m_c^2$, where $m_c$ is the mass of the charm quark.
At fixed target energies, where $Q^2$ is small, this 
approximation will break down but one can partially remedy this by also 
including threshold effects which are due to soft gluon bremsstrahlung.
The latter mechanism dominates the threshold region of heavy flavour 
production as is e.g. shown in \cite{mssn}, \cite{lsn1} 
for hadron-hadron scattering.
By including these threshold effects one can obtain a reasonable description
of the charm production cross section at small $Q^2$. 
We can show this for the unpolarized case since here the exact 
coefficient functions are available \cite{lrsn1}, \cite{rsn1}. 
Due to the similarity between the polarized and the
unpolarized cross sections in the threshold region
one can assume that this approximation will also 
work for $g_1(x,Q^2,m^2)$. In this way one can estimate the NLO effects
at the much smaller $Q^2$ values characteristic of fixed target 
experiments.

The content of the paper can be summarized as follows: in section 2 we
introduce our notations and give, for polarized Compton scattering, 
an exact analytic expression for the heavy quark coefficient 
function, which is valid for any $Q^2$ and $m^2$. 
In section 3 we compute the
full two-loop spin dependent operator matrix elements (OME's) 
contributing to the spin structure function $g_1(x,Q^2)$.
The spin dependent heavy quark coefficient functions will be presented 
in section 4 in the limit $Q^2 \gg m^2$. In section 5 we give 
improved expressions for them by
including threshold contributions so that they can be also used at smaller
$Q^2$ values. Furthermore we make estimates of the NLO corrections to polarized
charm production at HERA collider as well as fixed target energies.
The long formulae obtained for the full operator matrix elements and
the asymptotic heavy quark coefficient functions are presented
in appendices A and B respectively.   
 

\mysection{Heavy flavour production in polarized electron-proton scattering}

In this section we present the formulae needed to describe polarized heavy 
flavour electroproduction. Furthermore we summarize the findings 
in \cite{bmsmn}
how to derive the asymptotic form of the heavy quark coefficient functions
from the operator matrix elements and the light parton coefficient functions.
Heavy flavour production in polarized deep inelastic electron-proton scattering
proceeds via the following reaction
\begin{eqnarray}
e^-(\ell_1)+P(p)\rightarrow e^-(\ell_2)+Q(p_1)(\ \overline{Q}(p_2)\ )
+{}'X' \,.
\end{eqnarray}
Here $'X'$ represents any final inclusive hadronic state and the momenta 
of the heavy quark (anti-quark), denoted by $Q(\bar Q)$, are given 
by $p_1$ and $p_2$ respectively. The mass of the heavy quark $Q(\bar Q)$
is given by $m$.
Neglecting electro-weak radiative corrections the above process is dominated 
by the exchange of one vector boson which carries the 
momentum $q=\ell_1-\ell_2$. \\
If the virtuality of the exchanged vector boson $Q^2=-q^2>0$ is not too
large ( $Q^2\ll M_Z^2$ ) reaction $(2.1)$ proceeds via the exchange 
of one photon 
only. In this case the computation of the cross-section of $(2.1)$ 
involves the hadronic tensor
\begin{eqnarray}
W_{\mu \nu} (p,q,s) = {{1}\over{4 \pi}} \int \, d^4z \,e^{iq\cdot z}
< p,s |[ J_\mu(z), J_\nu(0)] |p,s> \,,
\end{eqnarray}
where $J_{\mu}$ stands for the electro-magnetic current and $s$ denotes
the spin vector of the proton with $s^2=-1$ and $s.p=0$.
The hadronic structure tensor can be split into symmetric and 
antisymmetric parts in the following way
\begin{eqnarray}
W_{\mu \nu} (p,q,s) = W_{\mu \nu}^S (p,q) +W_{\mu \nu}^A (p,q,s) \,,
\end{eqnarray}
\begin{eqnarray}
&&W_{\mu \nu}^S (p,q) =\frac{1}{2x}\Big(g_{\mu\nu}
-\frac{q_\mu q_\nu}{q^2}\Big)F_L(x,Q^2)
+\Big(p_\mu p_\nu-\frac{p.q}{q^2}(p_\mu q_\nu+p_\nu q_\mu)
\nonumber \\ && \qquad \qquad\qquad
+g_{\mu\nu}\frac{(p.q)^2}{q^2}\Big)\frac{F_2(x,Q^2)}{p.q}\,,
\end{eqnarray}
\begin{eqnarray}
W_{\mu \nu}^A (p,q,s)=-\frac{M}{2p.q}\epsilon_{\mu\nu\alpha\beta}
q^{\alpha}[s^{\beta}g_1(x,Q^2)+(s^{\beta}-\frac{s.q}{p.q}p^{\beta})
g_2(x,Q^2)]\,,
\end{eqnarray}
where $M$ is the mass of the proton and $x$ stands for the Bjorken scaling 
variable (see below).
The structure functions $F_L$ and $F_2$ already show up in unpolarized 
electron-proton scattering and the heavy flavour contributions have been 
extensively discussed up to next-to-leading order (NLO) in the strong 
coupling constant $\alpha_s$
in \cite{lrsn1}. If the incoming electron and proton are
polarized then in addition to $F_L$ and $F_2$ one also gets the
longitudinal spin structure function $g_1$ and the transverse spin 
structure function $g_2$. The latter show up in the polarized electron-proton
cross-section
\begin{eqnarray}
&&\frac{d^3\sigma^{\stackrel{\rightarrow}{\rightarrow}}}{dxdyd\phi}
-\frac{d^3\sigma^{\stackrel{\leftarrow}{\rightarrow}}}{dxdyd\phi}=
\frac{4\alpha^2}{Q^2}\Big[\Big\{2-y-\frac{2M^2x^2y^2}{Q^2}\Big\}g_1(x,Q^2)
\nonumber \\ && \qquad \qquad\qquad
-\frac{4M^2x^2y^2}{Q^2}g_2(x,Q^2)\Big] \,,
\end{eqnarray}
\begin{eqnarray}
&&\frac{d^3\sigma^{\stackrel{\uparrow}{\rightarrow}}}{dxdyd\phi}
-\frac{d^3\sigma^{\stackrel{\downarrow}{\rightarrow}}}{dxdyd\phi}=
-\frac{4\alpha^2}{Q^2}\cos \phi\Big(\frac{4M^2x^2}{Q^2}\Big)^{1/2}
\Big(1-y-\frac{M^2x^2y^2}{Q^2}\Big)^{1/2}
\nonumber \\ && \qquad \qquad\qquad
\times[yg_1(x,Q^2)+2g_2(x,Q^2)]\,.
\end{eqnarray}
The scaling variables $x$ and $y$ are defined by
\begin{eqnarray}
x=\frac{Q^2}{2p.q}\,,\quad (0<x<1)\,.\quad\quad y=\frac{p.q}{p.\ell_1}\,,\quad
(0<y<1)\,.
\end{eqnarray}
The angle between the spin $\vec s$ of the proton and the momentum 
$\vec \ell_2$
of the outgoing electron in $(2.1)$ is denoted by $\phi$.
The lower arrows on $\sigma$ in $(2.6), (2.7)$ indicate the polarization of
the incoming electron in the direction of its momentum $\vec \ell_1$. The
upper arrow on $\sigma$ in $(2.6)$ stands for the polarization of the proton
which is parallel or anti-parallel to the polarization of the incoming
lepton. The vertical arrows in $(2.7)$ also belong to the proton which is now 
polarized perpendicular (transverse) to the polarization of the lepton
in either the up or the down direction. In the subsequent part of this
paper we will limit ourselves to $g_1(x,Q^2)$ since it contains leading twist
two operators only whereas $g_2(x,Q^2)$ can also receive 
twist three contributions. In the case of twist two the 
structure functions can be described by the QCD-improved parton model.
In this model the heavy flavour contribution to $g_1$ can be expressed as
convolution integrals over the partonic scaling variable
$z=Q^2/(s+Q^2)$ where $s$ is the square of the photon-parton centre-of-mass 
energy $(s\ge 4m^2)$. This yields the following result
\begin{eqnarray}
 &&g_1(x,Q^2,m^2)=\frac{1}{2}\int^{z_{\rm max}}_x \frac{dz}{z}\Big[\frac{1}{n_f}
\sum^{n_f}_{k=1}e_k^2 \Big\{\Sigma\Big(\frac{x}{z},\mu^2\Big)
L^{\rm S}_q\Big(z,\frac{Q^2}{m^2},\frac{m^2}{\mu^2}\Big)
\nonumber \\ && \qquad\qquad
+G\Big(\frac{x}{z},\mu^2\Big)
 L_g^{\rm S}\Big(z,\frac{Q^2}{m^2},\frac{m^2}{\mu^2}
\Big)\Big\}
+\Delta\Big(\frac{x}{z},\mu^2\Big)
L^{\rm NS}_q\Big(z,\frac{Q^2}{m^2},\frac{m^2}{\mu^2}\Big)\Big]
\nonumber \\ && \qquad\qquad
+\frac{1}{2}e_Q^2\int^{z_{\rm max}}_{x}\frac{dz}{z}\Big\{
\Sigma\Big(\frac{x}{z},\mu^2\Big)
H_q^{\rm PS}\Big(z,\frac{Q^2}{m^2},\frac{m^2}{\mu^2}\Big)
\nonumber \\ && \qquad\qquad
+G\Big(\frac{x}{z},\mu^2\Big)
H_g^{\rm S}\Big(z,\frac{Q^2}{m^2},\frac{m^2}{\mu^2}\Big)
\Big\} \,,
\end{eqnarray}
where the upper boundary of the integration is given by
$z_{\rm max}=Q^2/(4m^2+Q^2)$. The function $G(z,\mu^2)$ stands for the polarized
gluon density. The singlet combination of the polarized quark densities is 
defined by
\begin{eqnarray}
\Sigma(z,\mu^2)=\sum^{n_f}_{i=1}\Big(f_i(z,\mu^2)+\overline{f}_i(z,\mu^2)
\Big) \,,
\end{eqnarray}
where $f_i$ and $\overline{f}_i$ stand for the light quark and anti-quark
densities of species $i$ respectively. The non-singlet combination of
the polarized quark densities is given by
\begin{eqnarray}
\Delta(z,\mu^2)=\sum^{n_f}_{i=1}\Big(e_i^2
-\frac{1}{n_f}\sum^{n_f}_{k=1}e_k^2\Big)
\Big(f_i(z,\mu^2)+\overline{f}_i(z,\mu^2)\Big)\,.
\end{eqnarray}
In the above expressions the charges of the light quark and the heavy quark
are denoted by $e_i$ and $e_Q$ respectively. Furthermore $n_f$ stands
for the number of light quarks and $\mu$ denotes the mass factorization
scale, which we choose to be equal to the  renormalization scale.
The latter shows up in the running coupling constant denoted by
$\alpha_s(\mu^2)$.

Like the parton densities the heavy quark coefficient
functions $L_i$ $(i=q,g)$ can also be divided 
into singlet and non-singlet parts, which are indicated by the
superscripts S and NS in $(2.9)$.
Furthermore the singlet quark coefficient function can be split into 
\begin{eqnarray}
L_q^{\rm S}  = L_q^{\rm NS}  + L_q^{\rm PS} \,.
\end{eqnarray}
The above relation originates from the light flavour decomposition
of the Feynman graphs contributing to the structure function
$g_1(x,Q^2,m^2)$ in (2.9).
One class of graphs gives the same contributions to $L_q^{\rm S}$
as well as to $L_q^{\rm NS}$ whereas another class, which we call 
purely-singlet (PS), only contributes to $L_q^{\rm S}$. The latter class
is characterized by those diagrams which have a gluon in the $t$-channel
and can therefore only contribute to the singlet quark coefficient function
$L_q^{\rm S}$. It turns out that up to order $\alpha_s^2$
$L_q^{\rm S} = L_q^{\rm NS} $ and $L_g^{\rm S} =0$.
In the case of the heavy quark coefficient functions $H_j$ there are no
non-singlet parts and therefore $H_q^{\rm S} = H_q^{\rm PS}$.
The latter gets contributions for the first time in order $\alpha_s^2$,
whereas the perturbation series in $H_g^{\rm S}$ starts in order $\alpha_s$.
The distinction between the heavy quark coefficient functions $L_i$ and 
$H_i$ can be traced back to the different photon-parton 
subprocesses from which they originate. The functions
$L_i$ are attributed to the reactions where the virtual photon
couples to the light quark, whereas $H_i$ originate from the reactions
where the virtual photon couples to the heavy quark. 
Hence $L_i$ and $H_i$ in (2.9) are multiplied by $e_i^2$
and $e_Q^2$ respectively. Moreover when the reaction where the
photon couples to the heavy quark contains a light quark in the initial
state then it can only proceed via the exchange of a gluon in the
$t$-channel between the light and heavy quark (see diagrams 5a, 5b
in \cite{lrsn1}). Therefore this process is
purely singlet and there does not exist a non-singlet contribution
to $H_q^{\rm S}$.

We will now discuss the various subprocesses which
contribute to the spin dependent heavy quark coefficient function up to
order $\alpha_s^2$. 
Both the coefficient functions $L_i$ and $H_i$ can be expanded in a power
series of $\alpha_s/(4\pi)$ with coefficients $L_i^{(n)}$, $H_i^{(n)}$ 
respectively, where $n$ denotes the order in the perturbation series.
Until now only the lowest order (LO) coefficient $H_g^{(1)}$ is exactly known
and it is given by the photon-gluon fusion process
\begin{eqnarray}
\gamma^*(q) + g(k_1) \rightarrow Q(p_1) + \overline{Q}(p_2) \,,
\end{eqnarray}
which yields the answer \cite{daw}, \cite{grv1}, \cite{v1}
\begin{eqnarray}
H_g^{(1)}(z,Q^2,m^2)=T_f\Big[4(2z-1)\ln\Big(\frac{1+sq_1}{1-sq_1}\Big)
+4(3-4z)sq_1\Big]
\end{eqnarray}
with 
\begin{eqnarray}
sq_1=\sqrt{1-\frac{4z}{(1-z)\xi}}\,,\quad
\xi = \frac{Q^2}{m^2}\,, \quad z = \frac{Q^2}{2 q\cdot k_1}\,.
\end{eqnarray}
where $0 < z < \xi/(\xi + 4)$ and the colour factor 
$T_f = 1/2$ in SU(N).
In NLO we encounter the following subprocesses. First we have the gluon 
bremsstrahlung subprocess
\begin{eqnarray}
\gamma^*(q) + g(k_1) \rightarrow g(k_2) + Q(p_1) + \overline{Q}(p_2) \,.
\end{eqnarray}
Including the virtual gluon corrections to $(2.13)$ we obtain the contribution
$H_g^{(2)}$ which is not yet known and for which the asymptotic form in the 
limit $Q^2\gg m^2$ will be determined in section 4.
In addition to $(2.16)$ we have the subprocess
\begin{eqnarray}
\gamma^*(q) +q(\overline{q})(k_1)\rightarrow q(\overline{q})(k_2)
+ Q(p_1) + \overline{Q}(p_2) \,,
\end{eqnarray}
which can be subdivided into two different production mechanisms.
The first one
is given by the Bethe-Heitler process (see figs. 5a,b in \cite{lrsn1})
and the second one can be attributed to 
the Compton reaction (see figs. 5c,d in \cite{lrsn1}).
In the case of the Bethe-Heitler process the virtual photon couples to
the heavy quark and therefore this reaction leads to $H_q^{(2)}$.
Like $H^{(2)}_g$ the coefficient function $H^{(2)}_q$ is not known 
and its asymptotic 
form will be  presented in section 4. In the Compton reaction the virtual 
photon couples to the light (anti-) quark and its contribution to
$L^{\rm NS}_q$ 
leads to $L^{{\rm NS},(2)}_q = L^{{\rm S},(2)}_q$ (see the discussion
below (2.12)).
Finally we want to make the comment that there are no interference
terms between the Bethe-Heitler and Compton reactions in (2.17)
if one integrates over all final state momenta. 

Like in the unpolarized case the computation
of the Compton process is rather trivial and we can present an exact
form of $L_q^{{\rm NS},(2)}$
\begin{eqnarray}
&& L_q^{{\rm NS},(2)}(z,\frac{Q^2}{m^2},\frac{m^2}{\mu^2})=C_FT_f\Big[\Big\{
\frac{4}{3}\frac{1+z^2}{1-z}-\frac{16 z}{1-z}\Big(\frac{z}{\xi}\Big)^2\Big\}
\Big\{\ln\Big(\frac{1-z}{z^2}\Big)L_1
\nonumber \\ && \qquad
+L_1L_2+2(-DIL1+DIL2+DIL3-DIL4)\Big\}
\nonumber \\ && \qquad
+\Big\{-\frac{8}{3}+\frac{4}{1-z}+\Big(\frac{z}{(1-z)\xi}\Big)^2\Big(
-16+32z-\frac{8}{1-z}\Big)\Big\}L_1
\nonumber \\ && \qquad
+\Big\{\frac{64}{9}+\frac{112}{9}z-\frac{152}{9}\frac{1}{1-z}
+\Big(\frac{z}{(1-z)\xi}\Big)\Big(\frac{512}{9}-\frac{128}{3}z
+\frac{848}{9}z^2\Big)
\nonumber \\ && \qquad
+\Big(\frac{z}{(1-z)\xi}\Big)^2\Big(-\frac{640}{9}+\frac{1408}{9}z
-\frac{2368}{9}z^2+\frac{1600}{9}z^3\Big)\Big\}\frac{L_3}{sq_2}
\nonumber \\ && \qquad
+\Big\{-\frac{188}{27}-\frac{872}{27}z+\frac{718}{27}\frac{1}{1-z}
+\Big(\frac{z}{(1-z)\xi}\Big)\Big(-\frac{952}{27}+\frac{1520}{27}z
\nonumber \\ && \qquad
-\frac{800}{9}z^2+\frac{20}{27}\frac{1}{1-z}\Big)\Big\}sq_1\Big]\,,
\end{eqnarray}
\noindent
where $sq_1$ is defined in (2.15) and $C_F = (N^2-1)/(2N)$ in SU(N).
Further we have defined
\begin{eqnarray}
&& \qquad \qquad sq_2=\sqrt{1-4\frac{z}{\xi}} \, ,
\nonumber \\ && 
L_1=\ln\Big(\frac{1+sq_1}{1-sq_1}\Big)
\quad , \quad L_2=\ln\Big(\frac{1+sq_2}{1-sq_2}\Big)
\quad , \quad L_3=\ln\Big(\frac{sq_2+sq_1}{sq_2-sq_1}\Big) \,,
\nonumber \\ && 
DIL_1={\rm Li}_2\Bigg(\frac{(1-z)(1+sq_1)}{1+sq_2}\Bigg)
\qquad , \qquad DIL_2={\rm Li}_2\Big(\frac{1-sq_2}{1+sq_1}\Big) \,,
\nonumber \\ && 
DIL_3={\rm Li}_2\Big(\frac{1-sq_1}{1+sq_2}\Big)
 \qquad , \qquad DIL_4={\rm Li}_2\Big(\frac{1+sq_1}{1+sq_2}\Big) \, .
\end{eqnarray}
Here Li${}_2(x)$ is the dilogarithm defined in \cite{lbmr}.

The derivation of the asymptotic formulae for the heavy quark coefficient 
functions has been given for the spin averaged structure functions $F_L$ and 
$F_2$ in \cite{bmsmn}. 
The procedure can be immediately carried over to the spin 
structure function $g_1$ and we only quote the results.

In the limit $Q^2\gg m^2$ the heavy quark coefficient functions $H_\ell$ 
can be obtained as follows
\begin{eqnarray}
H_{\ell}\Big(\frac{Q^2}{m^2},\frac{m^2}{\mu^2}\Big)=
A_{k\ell}\Big(\frac{m^2}{\mu^2}\Big)\otimes C_k\Big(\frac{Q^2}{\mu^2}\Big)\,,
\end{eqnarray}
with $k,\ell=q,g$ and $\otimes$ denotes the convolution symbol
\begin{eqnarray}
(f\otimes g)(z)=\int_0^1dz_1\int_0^1dz_2 \,\delta(z-z_1z_2)f(z_1)g(z_2)\,.
\end{eqnarray}
\noindent
Notice that for convenience we have suppressed the $z$-dependence in $(2.20)$.
The operator matrix elements (OME's) $A_{k\ell}$ are defined by
\begin{eqnarray}
A_{k\ell}\Big(\frac{m^2}{\mu^2}\Big)=<\ell|O_k|\ell>\,,
\end{eqnarray}
where $O_k$ $(k=q,g)$ are the composite quark and gluon operators
which appear in the operator product expansion (OPE) of the electro-magnetic
current in $(2.2)$ near the light-cone (see e.g. eq.(2.22) in \cite{bmsmn}).
In $(2.20)$ the operators are already renormalized and sandwiched between
on-shell light quark and gluon states indicated by $|\ell>$.
The latter leads to collinear divergences which, however, are removed via 
mass factorization so that the $A_{k\ell}$ are finite. Finally the OME's contain
one heavy quark loop only and they will be calculated up to $O(\alpha_s^2)$
in the next section.
The objects $C_k$ in $(2.20)$ denote the light quark and gluon coefficient
functions which contribute to the spin structure function $g_1$.
They have been calculated up to second order in \cite{zn}.
Notice that relation $(2.20)$ also holds for the heavy quark coefficient
functions of type $L_l$.

Finally we want to make the remark that, like in the case
of the coefficient functions, the OME's can also be divided
into non-singlet (NS), singlet (S) and purely-singlet (PS)
parts. In particular we have the relation
\begin{eqnarray}
A_{qq}^{\rm S} =  A_{qq}^{\rm NS} + A_{qq}^{\rm PS} \,,
\end{eqnarray}
and the origin of $A_{qq}^{\rm PS}$ is the same as mentioned below
(2.12) for the $L_q^{\rm PS}$ and $H_q^{\rm PS}$. Like in the
case of $H_q^{\rm PS}$ the OME's $A_{qq}^{\rm PS}$ are determined by the
graphs where two gluons are exchanged between the heavy quark and light quark
so that only the singlet channel can contribute.


\mysection{Calculation of the two-loop spin operator matrix elements}
In this section we present the calculation of the spin dependent OME's 
in $(2.22)$ containing 
one heavy quark loop up to two-loop order. A similar calculation
has been carried out for the spin averaged (unpolarized) OME's in \cite{bmsmn}.
The calculation of the spin dependent OME's proceeds in an analogous way.
The twist two operators $O_q$ and $O_g$ appearing in $(2.22)$ are
defined by  
\begin{eqnarray} 
O_{q,r}^{\mu_1\cdots\mu_n}(x)=i^nS\Big[\overline{\psi}(x)
\gamma_5\gamma^{\mu_1}D^{\mu_2}\cdots D^{\mu_n}\frac{\lambda_r}{2}\psi(x)\Big]
+{\rm trace\  terms} \,,
\end{eqnarray}
\begin{eqnarray} 
O_q^{\mu_1\cdots\mu_n}(x)=i^nS\Big[\overline{\psi}(x)
\gamma_5\gamma^{\mu_1}D^{\mu_2}\cdots D^{\mu_n}\psi(x)\Big]
+{\rm trace\  terms}  ~,
\end{eqnarray}
\begin{eqnarray} 
&&O_g^{\mu_1\cdots\mu_n}(x)=i^nS\Big[\frac{1}{2}\epsilon^{\mu_1\alpha\beta
\gamma}{\rm Tr}\Big( F_{\beta\gamma}(x)
D^{\mu_2}\cdots D^{\mu_{n-1}} F_{\alpha}^{\mu_n}(x)\Big)\Big]
\nonumber \\ && \qquad\qquad\qquad
+{\rm trace\  terms}  ~.
\end{eqnarray}
The symbol $S$ in front of the square bracket stands for the
symmetrization of the Lorentz indices $\mu_1\cdots\mu_n$ where $n$ denotes
the spin of the operator.
The above set of operators can be distinguished with respect to the flavour 
group $SU(n_f)$ in a non-singlet part $(3.1)$, carrying the flavour
group generator $\lambda_r$, and the singlet parts $(3.2)$ and $(3.3)$.  
The quark and the gluon field tensors are given by $\psi(x)$ and 
$F^a_{\mu\nu}(x)$ respectively with $F_{\mu\nu}=F_{\mu\nu}^a\,T^a$ where
$T^a$ stands for the generator of the colour group $SU(N)$ $(N=3)$.
The covariant derivative is denoted by 
$D_{\mu}=\partial_{\mu}+ig\,T^aA^a_{\mu}(x)$
where $A_{\mu}^a(x)$ represents the gluon field.
The operator vertices corresponding to the operators $(3.1)-(3.3)$ can
e.g. be found in appendix A of \cite{mn}. The heavy quark coefficient
functions mentioned in the last section require the calculation of the
following OME's (see (2.22)). Starting with the photon-gluon fusion
reaction $(2.13)$, $(2.16)$ we have to compute up to second order
\begin{eqnarray}
A_{Qg}^{\rm S,(n)}\Big({{m^2}\over{\mu^2}}\Big) = <g|O_Q^{(n)}(0)|g>  \,,
\end{eqnarray}
where $Q$ stands for the heavy quark with mass $m$. The OME's corresponding
to $(3.4)$ are determined by the Feynman graphs in fig. 1 (first order)
and fig. 2 (second order). For the Bethe-Heitler reaction $(2.17)$,
which starts in second order of $\alpha_s$, we need 
\begin{eqnarray}
A_{Qq}^{{\rm PS},(n)}\Big({{m^2}\over{\mu^2}}\Big) = <q|O_Q^{(n)}(0)|q>  \,.
\end{eqnarray}
Here the two-loop graphs are depicted in fig. 3. Finally we also want to 
obtain the asymptotic form of the coefficient function for the Compton reaction
$(2.17)$ presented in $(2.18)$. For this we need to calculate 
\begin{eqnarray}
A_{qq,Q}^{{\rm NS},(n)}\Big({{m^2}\over{\mu^2}}\Big) = <q|O_q^{(n)}(0)|q>  \,,
\end{eqnarray}
which is derived from the graphs in fig. 4
containing the heavy quark loop ($Q$) contribution to the
gluon self-energy.
Since the OME's $A_{k\ell}^{(n)}$ are S-matrix elements they originate from
the Fourier transform in momentum space of the following connected Green's
functions. The connected Green's function 
\begin{eqnarray}
<0|T(A_{\mu}^a(x)O_Q^{\mu_1\cdots\mu_n}(0)A_{\nu}^b(y)|0>_c\,,
\end{eqnarray}
corresponds to equation $(3.4)$. 
Here $O_Q^{\mu_1\cdots\mu_n}$ stands for the heavy quark composite operator
which is defined in the same way as in $(3.2)$ except that now the Dirac field
occurring in $O_Q^{\mu_1\cdots\mu_n}$ represents the heavy quark.
The heavy quark composite operator can be also sandwiched between
light quark fields denoted by $\psi_i(x)$
\begin{eqnarray}
<0|T(\overline{\psi}_i(x)O_Q^{\mu_1\cdots\mu_n}(0)\psi_j(y)|0>_c\,,
\end{eqnarray}
which correspond to $(3.5)$. Finally the connected Green's function
related to $(3.6)$ is given by
\begin{eqnarray}
<0|T(\overline{\psi}_i(x)O_{q,r}^{\mu_1\cdots\mu_n}(0)\psi_j(y)|0>_c\,,
\end{eqnarray}
where again $\psi_i(x)$ represents the light quarks and
$O_{q,r}^{\mu_1\cdots\mu_n}$ is the non-singlet light quark composite 
operator in $(3.1)$. In these connected Green's functions 
$a$ and $i,j$ are the colour indices of the gluon field $A_{\mu}$ and
the quark fields $\bar \psi$, $\psi$ respectively. 
Before one obtains the S-matrix elements
in $(3.4)$-$(3.6)$ the external gluon and quark propagators appearing
in $(3.7)$-$(3.9)$ have to be amputated. Further one has to include the 
external quark and gluon self-energies given by the graphs $s$, $t$ in
fig. 2. One can simplify the above 
connected Green's functions using the property that the composite operators 
are traceless symmetric tensors under the Lorentz group.
Therefore one obtains many trace terms which are not essential for the
determination of the OME's $A_{k\ell}^{(n)}$. These trace terms can be 
eliminated by multiplying the operators by an external source 
$J_{\mu_1\cdots\mu_n}=\Delta_{\mu_1}\cdots\Delta_{\mu_n}$ with $\Delta^2=0$.
Performing the Fourier transform into momentum space and sandwiching the 
connected Green's function $(3.7)$ by the external gluon polarization vectors
$\epsilon^{\mu}(p)$, $\epsilon^\nu(p)$, one obtains 
\begin{eqnarray}
\epsilon^{\mu}(p)G^{ab}_{Q,\mu\nu}\epsilon^{\nu}(p)\,,
\end{eqnarray}
where $p$ stands for the momentum of the external gluon in the graphs of figs.
1, 2 with $p^2=0$. The tensor $G_{Q,\mu\nu}^{ab}$ has the form
\begin{eqnarray}
G_{Q,\mu\nu}^{ab}=\hat{A}_{Qg}^{\rm S,(n)}\Big(\epsilon,
\frac{m^2}{\mu^2},\alpha_s\Big)
\delta^{ab}(\Delta.p)^{n-1}\epsilon_{\mu\nu\alpha\beta}\Delta^{\alpha}
p^{\beta}\,.
\end{eqnarray}
Proceeding in the same way for the connected Green's functions in $(3.8)$
and $(3.9)$ after multiplication by the Dirac
spinors $\bar u(\vec p,s)$, $u(\vec p,s)$ one obtains
\begin{eqnarray}
\overline{u}(\vec p,s)G_Q^{ij}u(\vec p,s)
\end{eqnarray}
and 
\begin{eqnarray}
\overline{u}(\vec p,s)G_q^{ij}\lambda_r u(\vec p,s)\,,
\end{eqnarray}
respectively. Here $\lambda_r$ denote the generators of the flavour group
$SU(n_f)$ and $u(\vec p,s)$ stands for the Dirac spinor corresponding 
to the light external quark in the
graphs of figs. 3, 4 with momentum $p$ $(p^2=0)$. The tensors $G_Q^{ij}$ and
$G_q^{ij}$ become equal to
\begin{eqnarray}
G_Q^{ij}=\hat{A}^{{PS},(n)}_{Qq}\Big(\epsilon, \frac{m^2}{\mu^2},\alpha_s\Big)
\delta^{ij}(\Delta.p)^{n-1}{\Delta \llap/}\gamma_5\,,
\end{eqnarray}
\begin{eqnarray}
G_q^{ij}=\hat{A}^{{NS},(n)}_{qq}\Big(\epsilon,
\frac{m^2}{\mu^2},\alpha_s\Big)
\delta^{ij}(\Delta.p)^{n-1}{\Delta \llap/}\gamma_5\,.
\end{eqnarray}
Notice that the operators appearing in the connected Green's functions
$(3.7)-(3.9)$ are still unrenormalized so that the OME's $\hat{A}_{k\ell}$
contain ultraviolet (UV) divergences. Besides the UV-singularities we also 
encounter collinear (C) divergences. They can be attributed to the
coupling of massless external quarks and gluons in the graphs of figs. 2-4
to internal massless quanta. Both types of singularities have to be 
regularized and we choose the method of $d$-dimensional regularization. 
This regularization
leads to pole terms of the type $1/\epsilon^i$ $(\epsilon=d-4)$ in the 
unrenormalized quantities $\hat{A}_{k\ell}$ which have to be removed via 
renormalization and mass factorization. Sometimes it is useful to distinguish
between UV- and C-divergences and, where appropriate, we will indicate them
by the notation $\epsilon_{UV}$ and $\epsilon_C$ respectively 
in the subsequent formulae. 
In general $d$-dimensional regularization is the
most suitable method to regularize all kinds of singularities appearing in 
perturbative quantum field theories since it preserves the Slavnov-Taylor
identities characteristic of gauge field theories. However, in the case
of spin quantities the $\gamma_5$-matrix appears together with the
Levi-Civita tensor (see eqs.(3.10)-(3.15)) for which one has to find a 
$d$-dimensional prescription. For the $\gamma_5$-matrix we will choose the 
prescription of 't Hooft and Veltman \cite{hv} which is equivalent to 
the one given by Breitenlohner and Maison in \cite{bm}.
Since in our calculations only one $\gamma_5$-matrix appears,
one can show  \cite{ad}, \cite{larin} that the replacement of 
${\Delta \llap/}\gamma_5$ in $(3.14)$, $(3.15)$ by
\begin{eqnarray}
{\Delta \llap/}\gamma_5=\frac{i}{6}\epsilon_{\mu\nu\rho\sigma}
\Delta^{\mu}\gamma^{\nu}\gamma^{\rho}\gamma^{\sigma}
\end{eqnarray}
is equivalent to the prescription given in \cite{hv}, \cite{bm}.
Although the methods in \cite{hv}-\cite{larin} are all consistent they 
have one drawback,
namely that the operator $O^{\mu}_{q,r}(3.1)$ gets renormalized in spite of 
the fact that it is conserved. Furthermore the $\gamma_5$-prescription
in $(3.16)$ also affects $O_{q,r}^{\mu_1\cdots\mu_n}$ ($n>1$) $(3.1)$
and $O_q^{\mu_1\cdots\mu_n}(3.2)$ so that one has to introduce some additional
renormalization constants to restore the Ward identities 
violated by $(3.16)$. Unfortunately these constants have been only calculated
in the literature \cite{larin} up to two-loop order for spin $n=1$ 
and we need the finite OME's $A^{(n)}_{k\ell}$ for arbitrary $n$.
However, as we will show later on, one can avoid the last procedure as long as
one combines quantities which are calculated using the same 
$\gamma_5$-prescription. This is what happens in the case of the heavy quark 
coefficient functions $H_{\ell}$ as presented in $(2.20)$.
If the OME's $A_{k\ell}$ and the light parton coefficient 
functions $C_k$
are computed using the same definition for the $\gamma_5$-matrix and
the Levi-Civita tensor the prescription dependence will cancel in the 
product on the right-hand side of $(2.20)$ providing us with a unique result
for $H_{\ell}$.

The unrenormalized $\hat{A}^{(n)}_{Qg}$ in $(3.11)$ can be obtained as follows.
First one replaces the $\gamma_5$-matrix appearing in $O_Q^{\mu_1\cdots\mu_n}$
in $(3.7)$ according to prescription $(3.16)$. Because of Lorentz covariance
the Levi-Civita tensor in $(3.11)$ emerges in a natural way.
This tensor can be projected out in $d$-dimensions using the relation
\begin{eqnarray}
\hat{A}_{Qg}^{\rm S,(n)}\Big(\epsilon,\frac{m^2}{\mu^2},\alpha_s\Big)=\frac{1}{N^2-1}
\frac{1}{(d-2)(d-3)}\epsilon^{\mu\nu\lambda\sigma}\delta^{ab}
(\Delta.p)^{-n}G^{ab}_{Q,\mu\nu}\Delta_{\lambda}p_{\sigma}\,.
\end{eqnarray}
In this way we obtain only Lorentz scalars in the numerators of the Feynman
integrals which can be partially cancelled by similar terms appearing in the 
denominators. The traces of the Feynman loops in figs. 1, 2 and the 
contractions over dummy Lorentz indices have been performed with the
algebraic manipulation program FORM \cite{jamv}.
The calculation of many of the scalar integrals has been 
already done in \cite{bmsmn}
for the spin averaged analogue of $\hat{A}^{(n)}_{Qg}$ and we can 
take over those results except for some additional integrals which
we have computed for the spin case in $(3.17)$.
In the discussion of our results we will drop the S in $A^{\rm S,(n)}_{Qg}$
and perform the inverse Mellin transformation on the OME's
so that they become dependent on the partonic variable $z$ in (2.15).
This implies that all the products are replaced by convolutions
according to $(2.21)$.

The one-loop OME $\hat{A}^{(1)}_{Qg}$, which only gets a non-zero contribution
from diagram $a$ in fig. 1, can be cast in the algebraic form
\begin{eqnarray}
\hat  A^{(1)}_{Qg}  = S_\epsilon \Big(\frac{m^2}{\mu^2}\Big)^{\epsilon/2}
\Big\{ - \frac{1}{\epsilon_{\rm UV}} P^{(0)}_{qg} + a^{(1)}_{Qg} +
\epsilon_{\rm UV} \bar a^{(1)}_{Qg} \Big\} \,.
\end{eqnarray}
Here $S_{\epsilon}$ is a spherical factor defined by
\begin{eqnarray}
S_{\epsilon}=\exp\Big\{\frac{\epsilon}{2}\Big[\gamma_E-\ln(4\pi)\Big]\Big\}\,,
\end{eqnarray}
which is characteristic of $d$-dimensional regularization and contains 
the Euler constant $\gamma_E$. The mass parameter $\mu$
originates from the dimensionality of the coupling constant $g$ 
($\alpha_s=g^2/(4\pi)$) in $d$-dimensions and should not be confused with the 
renormalization and mass factorization scales. However, if one only subtracts
the pole terms like in the $\overline{\rm MS}$-scheme, the mass parameter
$\mu$ turns into the afore-mentioned scales. The superscript $k$ in 
$\hat{A}^{(k)}_{ij}$ denotes the order in the perturbation series expansion of 
the OME's which can be written as
\begin{eqnarray}
\hat{A}_{ij} = \sum_{k=0}^{\infty} \Big( \frac{\alpha_s}{4\pi} \Big)^{k}
\hat{A}_{ij}^{(k)} \,.
\end{eqnarray}
The objects $P^{(k)}_{ij}$ ($i,j=q,g;k=0,1,\cdots$) which we will need
for $(3.18)$ and the subsequent expressions denote the spin AP-splitting 
functions which have been calculated up to next-to-leading order
in \cite{mn}, \cite{wv2}. In lowest order the renormalization group
coefficients in $(3.18)$ become
\begin{eqnarray}
&&P^{(0)}_{qg}=8T_f[2z-1]
\nonumber \\&&
a^{(1)}_{Qg}=0
\nonumber \\&&
\overline{a}^{(1)}_{Qg}=-\frac{1}{8}\zeta(2)P^{(0)}_{qg}\,.
\end{eqnarray}
The two-loop OME $\hat{A}^{(2)}_{Qg}$ which is determined by the graphs in
fig. 2 is given by $(A.1)$. As has been shown in section 3 of \cite{bmsmn},
it can be expressed into the renormalization group coefficients as follows
\begin{eqnarray}
&& \hat A_{Qg}^{(2)}  = S_\epsilon^2 \Big( \frac{m^2}{\mu^2}\Big)^\epsilon
\Big[ \frac{1}{\epsilon^2} \Big\{
\frac{1}{2} P^{(0)}_{qg} \otimes ( P^{(0)}_{qq} - P^{(0)}_{gg})
+ \beta_0 P^{(0)}_{qg} \Big\}
\nonumber  \\ \cr && \qquad 
+ \frac{1}{\epsilon} \Big\{ - \frac{1}{2} P^{(1)}_{qg}
-2 \beta_0 a^{(1)}_{Qg} - a^{(1)}_{Qg} \otimes (P^{(0)}_{qq} - P^{(0)}_{gg})
\Big\} + a^{(2)}_{Qg} \Big]
\nonumber  \\ \cr && \qquad
- \frac{2}{\epsilon} S_\epsilon\sum_{H=Q}^t \beta_{0,H}
\Big( \frac{m_H^2}{\mu^2} \Big)^{\epsilon/2}
\Big( 1 + \frac{1}{8}\epsilon^2 \zeta(2) \Big)
\hat A^{(1)}_{Qg} \, .
\end{eqnarray}
In the above expression, where mass renormalization has already been
carried out, the pole terms $\epsilon^{-k}$ stand for
the UV as well as C-divergences so that we have put 
$\epsilon_{UV}=\epsilon_C$.
The last term in $(3.22)$ can be traced back to the graphs $s$, $t$ in fig. 2.
which contain the heavy quark loop contributions to the external gluon
self-energy. Here one sums over all heavy quarks called $H$ starting with
$H=Q$ and ending with $H=t$ (top-quark). Notice that $Q$ is the lightest
heavy quark which is the same as the one produced in the final state of 
process $(2.1)$. Further it also appears in the heavy quark operator whose
OME's are shown in figs. 1, 2. For instance for charm production in $(2.1)$
we have $Q=c$, $m_c=m$ and the sum in $(3.22)$ runs over $H=c,b,t$.
In the case of bottom (b) production we have $Q=b$, $m_b=m$ and the sum
runs over $H=b,t$. The contribution to the beta-function from a
heavy quark $H$ is given by
\begin{eqnarray}
\beta_{0,H}=-\frac{4}{3}T_f \,,
\end{eqnarray}
for all species $H$ which implies that $\beta_{0,Q}=\beta_{0,H}$.
The Riemann zeta-function $\zeta(2)=\pi^2/6$ originates from the heavy
quark contribution to the gluon self-energy denoted by
$\Pi(p^2,m^2)$. At $p^2=0$ the unrenormalized expression
$\Pi(0,m^2)$ is proportional to $(1+\epsilon^2\zeta(2)/8)/\epsilon$.

The other renormalization group coefficients can be inferred from the
literature \cite{mn}, \cite{wv2}-\cite{sar} and they are given by
\begin{eqnarray}
&&\beta_0=\frac{11}{3}C_A-\frac{4}{3}n_fT_f \,,
\nonumber  \\ \cr && 
P^{(0)}_{qq}=4C_F\Big[2\Big(\frac{1}{1-z}\Big)_+ - 1 - z +
\frac{3}{2}\delta(1-z)\Big] \,,
\nonumber  \\ \cr && 
P^{(0)}_{gg}=8C_A\Big[\Big(\frac{1}{1-z}\Big)_++1-2z\Big]
+2\beta_0\delta(1-z) \,,
\nonumber  \\ \cr && 
P^{(1)}_{qg}=4C_FT_f\Big[2(1-2z)\Big(-2\ln^2(1-z)+4\ln z\ln(1-z)-\ln^2z
 +4\zeta(2)\Big)
\nonumber  \\ \cr && \qquad
  +16(1-z)\ln(1-z)-2(1-16z)\ln z+4+6z\Big]
\nonumber  \\ \cr && \qquad
 +4C_AT_f\Big[4(1-2z)\ln^2(1-z)-4(1+2z)\Big(\ln^2z+2\ln z\ln(1+z)
\nonumber  \\ \cr && \qquad
 +2{\rm Li}_2(-z)\Big)-8\zeta(2)+4(1+8z)\ln z-16(1-z)\ln(1-z)
\nonumber  \\ \cr && \qquad
+4(12-11z)
\Big]\,,
\end{eqnarray}
where the colour factor $C_A$ is given by $C_A=N$ in SU(N).

The full analytic expression for the unrenormalized 
$\hat A_{Qg}^{(2)}$ 
can be found in (A.1). Here the polylogarithmic functions
${\rm Li}_n(z)$ and ${\rm S}_{n,p}(z)$ are defined in \cite{lbmr}. 
In order to construct the
heavy quark coefficient functions in the next section we need to
renormalize $\hat A_{Qg}^{(\ell)}$ $(\ell = 0,1)$ in eqs. (3.18) and (3.22).
For the coupling constant renormalization we choose a scheme where the
heavy quarks $H$ appearing in the sum of $(3.22)$ decouple from the running
coupling constant $\alpha_s(\mu^2)$ when $\mu^2<m_H^2$. This renormalization 
prescription completely removes the last term in $(3.22)$ from the OME
$\hat A_{Qg}^{(2)}$. It also implies that only the light flavours
indicated by $n_f$ appear in the running coupling constant. For instance
in the case of charm production $n_f=3$ and the c, b and t quark
contributions are absent in the running coupling constant.

Furthermore one has to remove the C-divergences occurring in (3.22) via mass
factorization. The above procedure has been carried out in section 3 
of \cite{bmsmn}
in the context of the spin averaged OME and we can simply take over
the algebraic expressions from that paper. 
In the $\overline{\rm MS}$-scheme the renormalized OME's  
$ A_{Qg}^{(1)}$ and $ A_{Qg}^{(2)}$ become 
\begin{eqnarray}
 A_{Qg}^{(1)}  = - \frac{1}{2} P_{qg}^{(0)} \ln \frac{m^2}{\mu^2} 
+ a_{Qg}^{(1)} \,,
\end{eqnarray}
\begin{eqnarray}
&&  A_{Qg}^{(2)}  = \Big\{ \frac{1}{8} P_{qg}^{(0)}
\otimes( P_{qq}^{(0)} - P_{gg}^{(0)})
+\frac{1}{4} \beta_0 P_{qg}^{(0)} \Big\} \ln^2 \frac{m^2}{\mu^2} 
\nonumber  \\ \cr && \qquad
+\Big\{ - \frac{1}{2} P_{qg}^{(1)} - \beta_0 a_{Qg}^{(1)}
+\frac{1}{2} a_{Qg}^{(1)} \otimes 
( P_{gg}^{(0)} - P_{qq}^{(0)}) \Big\}
\ln \frac{m^2}{\mu^2}
\nonumber  \\ \cr && \qquad
+ a_{Qg}^{(2)} +2 \beta_0 \bar a_{Qg}^{(1)} + \bar a_{Qg}^{(1)} \otimes
( P_{qq}^{(0)} - P_{gg}^{(0)}) \,.
\end{eqnarray}
Notice that (3.26) is still dependent on the $\gamma_5$-matrix
prescription which enters the $C_F T_f$ part (see (A.1)).
This is exhibited by the splitting function $P_{qg}^{(1)}$ (3.24)
where the $C_F T_f$ part still has to undergo an additional renormalization
in order to become equal to the result in \cite{mn}, \cite{wv2}. 
The prescription
dependence also affects $a_{Qg}^{(2)}$.
However as will be shown in the next section this will be compensated
in the construction of the heavy quark coefficient functions when we add 
the same type of term coming from the massless parton coefficient function
$C_g^{(2)}$ computed in \cite{zn}.

Finally, the unrenormalized as well as renormalized OME's $A^{(\ell)}_{Qg}$
($\ell=1,2$) satisfy the relation
\begin{eqnarray}
\int_0^1dz\ A^{(\ell)}_{Qg}(z)=0\,.
\end{eqnarray}
The OME $A_{Qq}^{{\rm PS},(2)}$ is determined by the graphs in fig. 3. 
However it appears that only diagram $a$ gives a nonzero contribution.
The easiest way to compute this graph is by using the standard
Feynman parameterization which provides us with (see (3.14))
the unrenormalized OME 
\begin{eqnarray}
&&\hat A_{Qq}^{{\rm PS},(2)} = S_\epsilon^2\Big(\frac{m^2}{\mu^2}\Big)^\epsilon
\Big[ \frac{1}{\epsilon^2} \{ 
- \frac{1}{2} P_{qg}^{(0)} \otimes P_{gq}^{(0)} \} 
\nonumber  \\ \cr && \qquad
+ \frac{1}{\epsilon} 
 \{-\frac{1}{2} P_{qq}^{{\rm PS},(1)}  + a_{Qg}^{(1)} \otimes P_{gq}^{(0)} \}
+ a_{Qq}^{{\rm PS},(2)} \Big] \,. 
\end{eqnarray}
Like in the case of $\hat A_{Qg}^{(2)}$ (3.22) we did not make
any distinction between UV- and C-singular pole terms $\epsilon^{-k}$
$(\epsilon_{\rm UV} = \epsilon_{\rm C}$).
The renormalization group coefficients are given by
(see also (3.21))
\begin{eqnarray}
&& P_{gq}^{(0)} = 4 C_F [ 2 - z] \,, 
\nonumber  \\ \cr && 
P_{qq}^{{\rm PS},(1)} = 16 C_F T_f [ - (1+z) \ln^2 z - (1-3z)  \ln z
+ 1-z] \,.
\end{eqnarray}
The analytic expression for the unrenormalized $\hat A_{Qq}^{{\rm PS},(2)}$
can be found in (A.2). After removing the UV- and C-divergences
the renormalized OME reads (see section 3 of \cite{bmsmn})
\begin{eqnarray}
&&  A_{Qq}^{{\rm PS},(2)}  = \Big\{-\frac{1}{8} P_{qg}^{(0)}
\otimes P_{gq}^{(0)} \Big\} \ln^2 \frac{m^2}{\mu^2} 
\nonumber  \\ \cr && \qquad
+\Big\{ - \frac{1}{2} P_{qq}^{{\rm PS},(1)} 
+\frac{1}{2} a_{Qg}^{(1)} \otimes P_{gq}^{(0)} \Big\}
\ln \frac{m^2}{\mu^2}
\nonumber  \\ \cr && \qquad
+ a_{Qq}^{{\rm PS},(2)}  - \bar a_{Qg}^{(1)} \otimes
 P_{gq}^{(0)}  \,.
\end{eqnarray}
The above expression is represented in the $\overline{\rm MS}$-scheme.
Like $A_{Qg}^{(2)}$ in (3.26), 
$A_{Qq}^{{\rm PS},(2)}$ still depends on the prescription 
for the $\gamma_5$ matrix.
Since $P_{qq}^{{\rm PS},(1)}$ already agrees with the results in \cite{mn},
\cite{wv2}
it does not need any finite renormalization and the dependence on the
prescription for the $\gamma_5$ matrix only enters via the non-logarithmic
term $a_{Qq}^{{\rm PS},(2)}$.
While computing the heavy quark coefficient function 
the latter dependence will be cancelled 
by contributions from the massless quark coefficient function
$C_q^{{\rm PS},(2)}$ in \cite{zn}.

Finally we call attention to the non-singlet OME $A_{qq,Q}^{{\rm NS},(2)}$
derived from the graphs in fig. 4. These graphs are also computed
by Feynman parameterization and we get the unrenormalized result from (3.15) 
\begin{eqnarray}
&&\hat A_{qq,Q}^{{\rm NS},(2)}=S_\epsilon^2\Big(\frac{m^2}{\mu^2}\Big)^\epsilon
\Big[ \frac{1}{\epsilon^2_{\rm UV}} \Big\{
-\beta_{0,Q} P^{(0)}_{qq} \Big\}
\nonumber  \\ \cr && \qquad 
+ \frac{1}{\epsilon} \Big\{ - \frac{1}{2} P^{{\rm NS},(1)}_{qq,Q}
\Big\} + a^{{\rm NS}(2)}_{qq,Q} \Big] \,.
\end{eqnarray}
Notice that in fig. 4 the heavy flavour loop contribution to the gluon
self-energy, denoted by $\Pi (p^2, m^2)$, contains the heavy quark $Q$
with mass $m$ only. For the construction of the heavy quark coefficient
function $L_q^{(2)}$ we do not need the contribution of the other heavy
quarks (as mentioned below (3.22)) which have masses larger than $m$.
Contrary to $\hat A_{Qg}^{(2)}$ and
$\hat A_{Qq}^{{\rm PS},(2)}$ the above equation only contains
UV-divergences since the heavy quark $Q$ prevents 
$\hat A_{qq,Q}^{{\rm NS},(2)}$ from being C-singular provided that the gluon
self-energy is renormalized in such a way that 
$\Pi_{\rm R}(0, m^2) =0$.
Further in the non-singlet case we can completely remove the dependence
on the prescription for the $\gamma_5$-matrix since after renormalization
$( A_{qq}^{\rm NS})_{\rm spin}  =  (A_{qq}^{\rm NS})_{\rm spin\ ave}$.
This has to be so because the splitting function $P_{qq}^{\rm NS}$
is the same for the spin averaged and the spin dependent non-singlet operators
(see \cite{mn}, \cite{larin}, \cite{wv2}). 
Therefore after removing the $\gamma_5$ prescription dependence
$\hat A_{qq,Q}^{{\rm NS},(2)}$ becomes the same as in the spin
averaged case and we get
\begin{eqnarray}
&& \beta_{0,Q} = - \frac{4}{3} T_f  \,,
\nonumber  \\ \cr && 
P_{qq,Q}^{{\rm NS},(1)} = C_F T_f \Big[ - \frac{160}{9} 
\Big( \frac{1}{1-z}\Big)_+ + \frac{176}{9} z - \frac{16}{9}
\nonumber  \\ \cr && \qquad
- \frac{16}{3} \frac{1+z^2}{1-z} \ln z 
+ \delta(1-z) \Big( - \frac{4}{3} - \frac{32}{3} \zeta(2)\Big)\Big] \,.
\end{eqnarray}
The analytical expression for $\hat A_{qq,Q}^{{\rm NS},(2)}$ can be
found in (C.5), (C.6) of \cite{bmsmn}.
After renormalization the OME becomes
\begin{eqnarray}
&&  A_{qq,Q}^{{\rm NS},(2)}  = \Big\{-\frac{1}{4}\beta_{0,Q} P_{qq}^{(0)}
 \Big\} \ln^2 \frac{m^2}{\mu^2} 
+\Big\{ - \frac{1}{2} P_{qq,Q}^{{\rm NS},(1)} \Big\}
\ln \frac{m^2}{\mu^2}
\nonumber  \\ \cr && \qquad
+ a_{qq,Q}^{{\rm NS},(2)} + \frac{1}{4}\beta_{0,Q}\zeta(2) P_{qq}^{(0)} \,.
\end{eqnarray}

Summarizing our above results for the OME's we found that the coefficients
of the double and single pole terms can be inferred from the
AP splitting functions and the beta-function which are
published in the literature. In this way we have a check on the calculations
of $\hat A_{ij}$. The non-pole terms 
$a_{Qg}^{(2)}$ (3.26) $a_{Qq}^{{\rm PS},(2)}$ (3.30) 
and $a_{qq,Q}^{{\rm NS},(2)}$ (3.33) cannot be
predicted and are calculated in this paper. They are needed to compute the
order $\alpha^2_s$ spin-dependent heavy quark coefficient functions 
(2.20) up to the
non-logarithmic term which will be done in the next section.


\mysection{Heavy quark coefficient functions}
In this section we present the heavy quark coefficient functions
$H_\ell$ and $L_\ell$ $(\ell=q,g)$ defined in $(2.9)$ up to order
$\alpha_s^2$ in the asymptotic limit $Q^2\gg m^2$. In \cite{bmsmn}
we showed that, for the spin averaged case, the asymptotic limits 
for $H_\ell$ and $L_\ell$ can be expressed into the renormalization group
coefficients appearing in the algebraic expression for the OME's $A_{k\ell}$
and the massless parton coefficient functions $C_k$.
This derivation follows from the mass factorization theorem, which is also
applicable to the spin structure function $g_1(x,Q^2)$ in $(2.9)$.
The theorem states that the mass dependent terms of the type
$\ln^i\left({m^2}/{\mu^2}\right)\ln^j\left({Q^2}/{m^2}\right)$ 
occurring in the asymptotic 
expressions for $H_\ell$, $L_\ell$ can be factorized out in the following way
\begin{eqnarray}
H_\ell^{\rm S}\Big(\frac{Q^2}{m^2}, \frac{m^2}{\mu^2}\Big)
= A_{kl}^{\rm S}\Big(\frac{m^2}{\mu^2}\Big) \otimes C_k^{\rm S}
\Big(\frac{Q^2}{\mu^2}\Big)\,,
\end{eqnarray}
\begin{eqnarray}
L_\ell^r\Big(\frac{Q^2}{m^2}, \frac{m^2}{\mu^2}\Big)
= A_{kl}^r\Big(\frac{m^2}{\mu^2}\Big) \otimes C_k^r
\Big(\frac{Q^2}{\mu^2}\Big)\,,
\end{eqnarray}
where $H_\ell$, $L_\ell$ $(\ell=q,g)$ denote the spin dependent heavy quark
coefficient functions in the limit $Q^2\gg m^2$ and $r=$S, NS.
Up to order $\alpha^2_s$ the finite spin dependent OME's $A_{k\ell}$
$(k,\ell=q,g)$ have been calculated in section 3 and are given in Appendix A.
The spin dependent light parton coefficient functions $C_k$ have
been also calculated up to order $\alpha^2_s$ and they can be found in 
\cite{zn}. Notice that $A_{k\ell}$ as well as $C_k$ have been computed in the 
$\overline{\rm MS}$-scheme and in the case of $(4.1)$ both depend on the 
prescription for the $\gamma_5$-matrix. 
In the products appearing on the right hand
sides of $(4.1)$, $(4.2)$ the scheme dependence is only partially cancelled, 
which is revealed by the dependence of $H_\ell$, $L_\ell$ on the factorization
scale $\mu^2$. The latter originates from the coupling of a light parton 
(gluon or quark) to an internal light parton which appears in
subprocesses $(2.16)$, $(2.17)$.
Fortunately the $\gamma_5$-matrix prescription cancels in the product
on the right hand side of $(4.1)$ as we will discuss below.
Since the algebraic structure of the spin dependent heavy quark
coefficient functions corresponding to $g_1(x,Q^2)$ in $(2.9)$
is the same as derived for the spin averaged structure function 
$F_2(x,Q^2)$ we can simply take over the formulae in \cite{bmsmn}.
For the LO  photon-gluon fusion process $(2.13)$ we obtain
\begin{eqnarray}
H^{(1)}_g \Big(\frac{Q^2}{m^2},\frac{m^2}{\mu^2}\Big) = 
\frac{1}{2} P^{(0)}_{qg}\ln\frac{Q^2}{m^2} + a^{(1)}_{Qg} 
+ c^{(1)}_g \,,
\end{eqnarray}
where $P^{(0)}_{qg}$ and $a^{(1)}_{Qg}$ appear in $A^{(1)}_{Qg}$ $(3.25)$
and $c^{(1)}_g$ is the non-log term in the lowest order coefficient
function defined by (see \cite{zn})
\begin{eqnarray}
C^{(1)}_g \Big(\frac{Q^2}{\mu^2}\Big) = 
\frac{1}{2} P^{(0)}_{qg}\ln \frac{Q^2}{\mu^2} + c^{(1)}_g \,.
\end{eqnarray}
Notice that like $A_{k\ell}$ in $(3.20)$ the coefficient functions are
expanded as 
\begin{eqnarray}
C_k\Big(\frac{Q^2}{\mu^2}\Big) = \sum_{\ell=0}^{\infty} 
\Big( \frac{\alpha_s}{4\pi} \Big)^{\ell}
C_k^{(\ell)}\Big(\frac{Q^2}{\mu^2}\Big) \,.
\end{eqnarray}
The explicit expression for $(4.3)$ can be found in $(B.1)$.
In NLO the spin dependent heavy quark coefficient
function of the photon-gluon fusion process $(2.13)$, $(2.16)$ becomes
\begin{eqnarray}
&& H^{(2)}_g\Big(\frac{Q^2}{m^2}, \frac{m^2}{\mu^2}\Big) = 
\Big\{ \frac{1}{8} P^{(0)}_{qg}\otimes( P^{(0)}_{gg} + P^{(0)}_{qq})
- \frac{1}{4} \beta_0 P^{(0)}_{qg}\Big\} \ln^2 \frac{Q^2}{m^2}
\nonumber \cr && \qquad
+ \Big\{  \frac{1}{2} P^{(1)}_{qg} - \beta_0 c^{(1)}_{g}
+ \frac{1}{2} P^{(0)}_{qq}\otimes a^{(1)}_{Qg}
+ \frac{1}{2} P^{(0)}_{gg}\otimes c^{(1)}_{g} 
+ \frac{1}{2} P^{(0)}_{qg}\otimes c^{(1)}_{q} 
\Big\} 
\nonumber \cr && \qquad 
\times \ln\frac{Q^2}{m^2}
+\Big\{ \frac{1}{4} P^{(0)}_{qg} \otimes P^{(0)}_{gg} 
- \frac{1}{2} \beta_0 P^{(0)}_{qg}\Big\} \ln\frac{Q^2}{m^2} 
\ln \frac{m^2}{\mu^2}
\nonumber \cr && \qquad 
+\Big\{ -\beta_0( c^{(1)}_{g} + a^{(1)}_{Qg})
+ \frac{1}{2} P^{(0)}_{gg} \otimes ( c^{(1)}_{g} + a^{(1)}_{Qg})
\Big\} \ln\frac{m^2}{\mu^2} 
\nonumber \cr && \qquad
+ c^{(2)}_{g} + a^{(2)}_{Qg}
+ 2 \beta_0 \bar a^{(1)}_{Qg}
+ c^{(1)}_{q} \otimes  a^{(1)}_{Qg}
+ P^{(0)}_{qq} \otimes  \bar a^{(1)}_{Qg}
- P^{(0)}_{gg} \otimes  \bar a^{(1)}_{Qg} \,,
\nonumber \\  
\end{eqnarray}
where the coefficients $P^{(1)}_{qg}$, $P^{(0)}_{ij}$, $a^{(\ell)}_{Qg}$,
$\bar a^{(\ell)}_{Qg}$ show up in $A^{(2)}_{Qg}$ $(3.26)$.
The coefficients $c^{(\ell)}_g$ $(\ell=1,2)$ show up in the order $\alpha^2_s$
contribution to the gluon coefficient function calculated in \cite{zn}.
The latter can be written as
\begin{eqnarray}
&& C^{(2)}_g\Big(\frac{Q^2}{\mu^2}\Big) = 
\Big\{ \frac{1}{8} P^{(0)}_{qg}\otimes( P^{(0)}_{gg} + P^{(0)}_{qq})
- \frac{1}{4} \beta_0 P^{(0)}_{qg}\Big\} \ln^2 \frac{Q^2}{\mu^2}
\nonumber \cr && \qquad \qquad
+ \Big\{  \frac{1}{2} P^{(1)}_{qg} - \beta_0 c^{(1)}_{g}
+ \frac{1}{2} P^{(0)}_{gg}\otimes c^{(1)}_{g}
+ \frac{1}{2} P^{(0)}_{qg}\otimes c^{(1)}_{q} \Big\}
\ln\frac{Q^2}{\mu^2}
\nonumber \\  && \qquad \qquad
+ c^{(2)}_{g}  \,.
\end{eqnarray}
The explicit expression for $(4.6)$ is given in $(B.2)$.
The latter can be split into $C_FT_f$ and $C_AT_f$ parts (see (B.2)).
The $C_AT_f$ part still depends on the mass factorization scale
$\mu^2$ and is therefore scheme dependent (in our case the 
$\overline{\rm MS}$-scheme).

The same scale dependence was also found for the exact expression of
$H_g^{(2)}$  (4.6) in \cite{lrsn1}.
It can be attributed to the coupling constant renormalization
represented by the lowest order coefficient $\beta_0$ in the 
beta-function and to mass factorization which is revealed by the lowest order
splitting function $P^{(0)}_{gg}$. The $C_F T_f$ part of $H_g^{(2)}$
is scheme independent which implies that it is obtained without performing
renormalization and mass factorization on the original parton cross
section of the photon-gluon fusion process $(2.13)$, $(2.16)$.
Therefore the latter did not contain UV and C-divergences and it can be
computed in four dimensions. From section 3 and \cite{zn}
we infer that the prescription dependence of the $\gamma_5$-matrix
only enters in the $C_F T_f$ parts of $A_{Qg}^{(2)}$ (3.26)
and $C_g^{(2)}$ (4.7).
Hence it has to cancel in the same part of $H_g^{(2)}$ because the latter
can be computed in four dimensions, where the
$\gamma_5$-matrix is unique. Further we made an interesting observation
for the gluonic coefficient functions $H_g$ and $C_g$ (see \cite{zn}) i.e,
\begin{eqnarray}
\int_0^1 \, dz\ H_g^{(\ell)} ( z, \frac{Q^2}{m^2}, \frac{m^2}{\mu^2})
=0 \,, \qquad 
\int_0^1 \, dz\ C_g^{(\ell)} ( z, \frac{Q^2}{\mu^2}) =0\,,
\end{eqnarray}
with $\ell = 1,2$.
This property was already mentioned for $A_{Qg}^{(\ell)}$
(see (3.27)).

The asymptotic expression of the order $\alpha_s^2$ spin dependent heavy
quark coefficient function corresponding to the Bethe-Heitler process
in $(2.17)$ is given by
\begin{eqnarray}
&&H_q^{(2)}( \frac{Q^2}{m^2} , \frac{m^2}{\mu^2}) =
\Big\{ \frac{1}{8} P_{qg}^{(0)} \otimes P_{gq}^{(0)} \Big\}
\ln^2 \frac{Q^2}{m^2}
+ \Big\{ \frac{1}{2} P_{qq}^{{\rm PS},(1)} +
\frac{1}{2} P_{gq}^{(0)} \otimes c_g^{(1)}\Big\} \ln \frac{Q^2}{m^2}
\nonumber \\  && \qquad \qquad
+ \Big\{ \frac{1}{4} P_{qg}^{(0)} \otimes P_{gq}^{(0)} \Big\}
\ln \frac{Q^2}{m^2} \ln \frac{m^2}{\mu^2}
+ \Big\{ \frac{1}{2} P_{gq}^{(0)} \otimes (c_g^{(1)} + a_{Qg}^{(1)})\Big\}
\ln \frac{m^2}{\mu^2}
\nonumber \\  && \qquad \qquad
+ c_q^{{\rm PS},(2)}  
+ a_{Qq}^{{\rm PS},(2)}  
- P_{gq}^{(0)}\otimes \bar a_{Qg}^{(1)}  \,. 
\end{eqnarray}
The coefficients $P_{qq}^{{\rm PS},(1)}$\,,\, $P_{ij}^{(0)}$\,,\,
$a_{Qq}^{{\rm PS},(2)}$  and $ \bar a_{Qg}^{(1)}$
show up in $A_{Qq}^{{\rm PS},(2)}$ (3.30)
and $C_{q}^{{\rm PS},(2)}$ appears in the pure singlet quark coefficient
function calculated in \cite{zn}
\begin{eqnarray}
&& C_q^{{\rm PS},(2)}\Big( \frac{Q^2}{\mu^2}\Big) =
\Big\{ \frac{1}{8} P_{qg}^{(0)} \otimes P_{gq}^{(0)} \Big\}
\ln^2 \frac{Q^2}{\mu^2}
\nonumber \\  && \qquad \qquad
+ \Big\{ \frac{1}{2} P_{qq}^{{\rm PS},(1)} +
\frac{1}{2} P_{gq}^{(0)} \otimes c_g^{(1)} \Big\} \ln \frac{Q^2}{\mu^2}
+ c_q^{{\rm PS},(2)} \,.
\end{eqnarray}
The explicit expression for (4.9) can be found in (B.3). Notice
that $H_q^{(2)}$ is still scheme dependent (in our case the
$\overline{\rm MS}$ scheme), which is indicated by the mass 
factorization scale $\mu^2$. This factorization scheme dependence 
shows up via the splitting function $P_{gq}^{(0)}$.
The $\gamma_5$-matrix prescription dependence enters via the 
nonlogarithmic terms in $A_{Qq}^{{\rm PS},(2)}$ (3.30) as well as in
$C_q^{{\rm PS},(2)}$ (4.10) and therefore cancels in the sum of both
which equals $H_q^{(2)}$.

Finally we turn our attention to the Compton process in (2.17) which provides
us with the spin dependent coefficient function $L_q^{\rm NS}$.
As can be inferred from (2.18) it is scheme independent since there is
no dependence on $\mu^2$. Furthermore the prescription dependence
for the $\gamma_5$-matrix could be removed from
$A_{qq,Q}^{{\rm NS},(2)}$ as well as from the order $\alpha_s^2$
non-singlet coefficient function $C_{q,Q}^{{\rm NS},(2)}$.
The latter reads (see (4.27) in \cite{bmsmn})
\begin{eqnarray}
&& C_{q,Q}^{{\rm NS},(2)}\Big( \frac{Q^2}{\mu^2}, \frac{m^2}{\mu^2}\Big) = 
\Big\{ - \frac{1}{4} \beta_{0,Q} P_{qq}^{(0)} \Big\}
\ln^2 \frac{Q^2}{m^2}
\nonumber \\  && \qquad \qquad
+ \Big\{  \frac{1}{4} \beta_{0,Q} P_{qq}^{(0)} \Big\}
\ln^2 \frac{m^2}{\mu^2}
+ \frac{1}{2} P_{qq,Q}^{{\rm NS},(1)}  \ln \frac{Q^2}{\mu^2}
\nonumber \\  && \qquad \qquad
-\beta_{0,Q} c_q^{(1)} \ln\frac{Q^2}{m^2}
+ c_{q,Q}^{{\rm NS},(2)} \,.
\end{eqnarray}
In the above expression we have chosen the same renormalization for the 
heavy quark loop ($Q$) contribution to the gluon self-energy as the one 
appearing in the 
OME  $A_{qq,Q}^{{\rm NS},(2)}$ (see below (3.31)).
Finally the heavy quark coefficient function is given by
\begin{eqnarray}
&& L_q^{{\rm NS},(2)}\Big( \frac{Q^2}{\mu^2}, \frac{m^2}{\mu^2}\Big) = 
\Big\{ - \frac{1}{4} \beta_{0,Q} P_{qq}^{(0)} \Big\}
\ln^2 \frac{Q^2}{m^2}
\nonumber \\  && \qquad \qquad
+ \Big\{ \frac{1}{2} P_{qq,Q}^{{\rm NS},(1)}- \beta_{0,Q} c_q^{(1)} \Big\}
\ln \frac{Q^2}{m^2}
+ c_{q,Q}^{{\rm NS},(2)} + a_{qq,Q}^{{\rm NS},(2)} 
\nonumber \\  && \qquad \qquad
+ \frac{1}{4} \beta_{0,Q} \zeta(2) P_{qq}^{(0)} \,,
\end{eqnarray}
where the coefficients 
$P_{qq,Q}^{{\rm NS},(1)}$\,,\, $P_{qq}^{(0)}$\,,\,
$\beta_{0,Q}$ and $a_{qq,Q}^{{\rm NS},(2)}$
show up in $A_{qq,Q}^{{\rm NS},(2)}$ (3.33). 
The coefficient $c_{q,Q}^{{\rm NS},(2)}$ can be found in \cite{zn}.
The explicit expression for (4.12) can be found in (B.4).
One can check that in the limit $Q^2 \gg m^2$ the exact expression for the
Compton process in (2.18) becomes equal to the asymptotic formula in (4.12)
or in (B.4).


\mysection{Results}
In this section we compute the heavy charm component of the spin structure
function $g_1(x,Q^2)$ and compare it with the light parton contributions.
As has been mentioned before (see section 2) the order $\alpha_s$ spin
dependent heavy quark coefficient function $H^{(1)}_g$ (2.14) is exactly 
known whereas the order $\alpha_s^2$ coefficient functions computed
in section 4 are only valid when $Q^2\gg m^2$. Therefore the latter can 
only be used at those $Q^2$-values which are characteristic 
for the polarized electron
and proton beam facility at HERA \cite{vwh}, \cite{jb}.
If we want to use the asymptotic expressions at the lower $Q^2$-values,
which are typical for the SMC-experiment \cite{gkm}, we have to improve them.
To this purpose we will construct below some improved 
coefficient functions which can be also used at smaller
$Q^2$-values. To check the validity of this approximation we carry out
the same procedure for  the spin averaged structure function
$F_2(x,Q^2,m^2)$. The latter is expressed in parton densities
and coefficient functions in the same way as done for $g_1(x,Q^2,m^2)$
in (2.9). The heavy quark coefficient functions needed for both $g_1$ and $F_2$
start in the same order of $\alpha_s$ and they have the same asymptotic
behaviour exhibited by the large logarithmic
terms $\ln^i(m^2/\mu^2)\ln^j(Q^2/m^2)$. Further the spin averaged as well as 
the spin dependent $H^{(2)}_g$ show the same threshold behaviour which is 
determined by soft gluon bremsstrahlung \cite{mssn}, \cite{lsn1}.
However contrary to $g_1$ the exact heavy quark coefficient 
functions are known for $F_2$ and are published in \cite{lrsn1}.
Therefore we can test the quality of the approximation for $F_2$ and
assume that it also works for $g_1$ using the same values for $Q^2$.
Since we already have the exact form of the coefficient functions
$L^{\rm NS,(2)}_q$ corresponding to the Compton process in (2.17) 
there is no need for an approximation here. 
The same holds for the Born approximation given by
$H_g^{(1)}$ in (2.14). 

In the case of the bremsstrahlung
reaction (2.16) and the Bethe-Heitler process (2.17), which lead to the 
heavy quark coefficient functions $H^{(2)}_g$ and $H^{(2)}_q$
respectively, we propose the following approximation
\begin{eqnarray}
&& H^{(2),\rm approx}_g\Big(z,\xi,\frac{m^2}{\mu^2}\Big)=
\Big(1-\frac{4m^2}{s}\Big)^{1/2}H^{(2)}_g\Big(z,\xi,\frac{m^2}{\mu^2}\Big)
\nonumber \\ && \qquad
+\Big(\frac{4m^2}{s}\Big)^{1/4}H^{(1)}_g\Big(z,\xi\Big)
S_{\rm thres}(s,m^2)
\end{eqnarray} 
with 
\begin{eqnarray}
&&S_{\rm thres}(s,m^2)=C_A\Big[4\ln^2\Big(1-\frac{4m^2}{s}\Big)
-20\ln\Big(1-\frac{4m^2}{s}\Big) 
\nonumber \\ && \qquad
+ 4 \ln\frac{m^2}{\mu^2}
\ln\Big(1-\frac{4m^2}{s}\Big)\Big]
\nonumber \\ && \qquad
+\Big(C_F-\frac{C_A}{2}\Big)\frac{2\pi^2}{\sqrt{1-4m^2/s}}
\end{eqnarray} 
and 
\begin{eqnarray}
&& H^{(2),\rm approx}_q\Big(z,\xi,\frac{m^2}{\mu^2}\Big)=
\Big(1-\frac{4m^2}{s}\Big)^{1/2}H^{(2)}_q\Big(z,\xi,\frac{m^2}{\mu^2}\Big)\,,
\end{eqnarray}
where $\xi$ and $z$ are defined in (2.15) and $s$ is the virtual photon-parton
CM energy. Expressed in $z$ and $\xi$ the latter becomes equal to
\begin{eqnarray}
s=\Big(\frac{1-z}{z}\Big)\xi m^2\,.
\end{eqnarray} 
The spin dependent heavy quark coefficient functions on the 
right hand side of (5.1), (5.3) are given in appendix B and they are
strictly valid for $\xi=Q^2/m^2\gg 1$. To improve their behaviour near 
threshold ($s=4m^2$) we multiply them by the factor $(1-4m^2/s)^{1/2}$.
Further we add to (5.1) a term which is obtained by multiplying the exact Born
coefficient function $H^{(1)}_g$ (2.14) by the factor $S_{\rm thres}$ (5.2).
The logarithmic terms in this factor originate from soft gluon bremsstrahlung
which is the dominant production mechanism near the threshold in process
(2.16) (see \cite{mssn}, \cite{lsn1}).
The last term in (5.2) represents the Coulomb singularity which can be 
attributed to the loop-graphs where one gluon is exchanged between the massive 
quark antiquark pair. 
The factor $S_{\rm thres}$ is universal (see \cite{lsn1}) and is the
same for $F_2$ and $g_1$. It has been computed for $F_2$ in eq. (5.7) of
\cite{lrsn1}. To suppress unwanted effects at larger $s$ we have removed
the factor 8 in the argument of the logarithms in eq. (5.7) of \cite{lrsn1}
and multiplied $S_{\rm thres}$ with $(4m^2/s)^{1/4}$ so that the second
part in (5.1) vanishes when $s \gg m^2$.

To test the approximation made above we apply it first to $F_2(x,Q^2,m^2)$
for which the exact \cite{lrsn1} as well as the asymptotic coefficient 
functions $H^{(2)}_{2,l}$ ($l=q,g$), valid for $Q^2\gg m^2$, (see appendix D
in \cite{bmsmn}) are known. To that purpose we plot the ratio
\begin{eqnarray}
R(x,Q^2,m^2)=\frac{F_2^{\rm approx}(x,Q^2,m^2)}{F_2^{\rm exact}(x,Q^2,m^2)}
\end{eqnarray}
in NLO and examine below for which $\xi$ (or $Q^2$) this 
approximation breaks down.
Here $F_2$ is given by the same  formula as in eq. (2.9) where now the heavy 
quark coefficient functions stand for the spin averaged ones. In eq. (5.4)
the superscripts 'exact' and 'approx' indicate whether the exact heavy quark 
coefficients in \cite{lrsn1} are used or the spin averaged
analogues of the approximations in (5.1), (5.3).
Notice that for the Born coefficient functions the exact formula
has been taken.

For our plots of $R$ in (5.5) we have chosen the parton density set GRV in 
the $\overline{\rm MS}$-scheme \cite{grv2}. Further we limit ourselves to charm
production ($m_c=m=1.5$ GeV${}/c^2$ ) which implies that the number of light 
flavours $n_f$ in (2.9) has to be taken to be three in the coefficient 
functions and the running coupling constant $\alpha_s(\mu^2)$ 
($\Lambda^{(3)}_{\bar{\rm MS}}=248$ MeV). 
The factorization/renormalization scale 
$\mu^2$ is set equal to $Q^2$. We have studied 
$R(x,Q^2,m^2)$ in the range $10^{-4}<x<1$ and $10<Q^2<10^4$ $({\rm GeV}/c)^2$. 

In fig. 5 we have plotted $R$ (5.5) as a function of $Q^2$ at 
$x=0.1,0.01,10^{-3}$ and $10^{-4}$. From this figure one infers that for
$Q^2 > Q^2_{\rm min} >20$ $({\rm GeV}/c)^2$  $R$ is very close to one
(actually $0.9<R<1.1$) which means that above this value $F_2^{\rm exact}$
and $F_2^{\rm approx}$ coincide. Even for $Q^2 > 10$ $({\rm GeV}/c)^2$
the approximation is rather good if one bears in mind that the statistics
of deep inelastic charm production is quite low. For $x>0.1$ the 
approximation gets worse which is revealed by fig. 6. Here $R>1.2$ which 
happens for $x=0.2$ when $Q^2=10$ $({\rm GeV}/c)^2$ or $x>0.5$ when $Q^2=100$
$({\rm GeV}/c)^2$. Therefore we can conclude that the approximation breaks down
at large $x$ and small $Q^2$-values. In the case of the HERA collider this is
not bad because the large $x$-regime is not accessible. However for fixed
target experiments where $x$ is large and $Q^2$ is small our predictions of the
NLO corrections have to be considered as an order-of-magnitude estimate only.

Next we study the validity of our approximation for the charm component of
the spin structure function denoted by $g_1(x,Q^2,m^2)$ (2.9). Here
we choose the leading log (LL) parametrization and the next-to-leading
log (NLL) parametrization in the $\overline{\rm MS}$-scheme for the
spin parton densities in \cite{grsv}. Here one has two sets of parton
densities which are obtained in the standard scenario and the valence
scenario. Our subsequent plots are made in the standard scenario although we
also studied the valence one. It turns out that the differences between
both scenarios are irrelevant for charm production so that we do not present
separate figures for the valence scenario.
Further the number of flavours and the running coupling constant are the same
as those taken for $F_2(x,Q^2,m^2)$ above. Since the exact coefficient
functions $H_l^{(2),\rm exact}$ ($l=q,g$) are unknown a comparison between the
exact and approximate spin structure function $g_1$ can be only made on
the Born level. Furthermore $g_1$ is not a positive definite quantity
either in the exact or approximate formulae. This implies that the numerator
and the denominator in (5.5) can vanish so that it makes no sense to plot
$R$ in the polarized case. Therefore we have to study the approximation
on the level of the structure function itself.

In figs. 7a,7b and 7c we have plotted $g_1^{\rm exact}$ and $g_1^{\rm approx}$
on the Born level for  $Q^2=10, 50$ and $100$ $({\rm GeV}/c)^2$
respectively. Here $g_1^{\rm approx}$ is obtained from 
the asymptotic coefficient function in 
(B.1) by multiplying the latter with $(1-4m^2/s)^{1/2}$. At $Q^2=10$ 
$({\rm GeV}/c)^2$ (fig. 7a) the deviation between the exact and approximated
 result is
of the order of $25 \%$ for $x<0.02$. For larger $x$-values the approximation
becomes much better. When  $Q^2$ increases (see figs. 7b,7c) $g_1^{\rm exact}$
(Born)
and $g_1^{\rm approx}$(Born) almost coincide over the whole $x$-region.
In NLO we have to take the approximate order $\alpha_s$ correction to
the charm structure function $g_1(x,Q^2,m^2)$ because not all exact
order $\alpha_s^2$ heavy quark coefficient functions are known. Denoting
this approximation by $g_1^{\rm approx}$(NLO) we assume that its validity
holds for the same $x$ and $Q^2$-values as those observed for 
$F_2^{\rm approx}$ above i.e. $x < 0.1$ and $Q^2 \gg 10$ $({\rm GeV}/c)^2$.
 For $x > 0.1$ we expect that the exact NLO charm structure function will
be very small which is already indicated by the Born contribution in 
figs. 7a,7b and 7c. The expression for $g_1^{\rm approx}$(NLO) is obtained
by adding to $g_1^{\rm exact}$(Born)  the order $\alpha_s$ corrections.
The latter originate from the exact expression for $L_q^{\rm NS,(2)}$
(2.18) and $H_l^{(2),\rm approx}$ ($l=q,g$) given in (5.1), (5.3).
The results for $g_1^{\rm approx}$ 
up to NLO are presented in figs. 8a, 8b and 8c for $Q^2=10, 50$ and $100$ 
$({\rm GeV}/c)^2$ respectively,
where they are compared with the Born contribution. From these figures we
infer that the bulk of the correction occurs in the region $0.01<x<0.1$
and amounts to almost $100 \%$ ( $g_1^{\rm approx}/g_1^{\rm exact}
({\rm Born}) \sim 2)$.

In figs. 9a, 9b and 9c we have plotted in NLO $g_1^{\rm approx}(x,Q^2,m^2)$ and 
$g_1^{\rm light}(x,Q^2)$ where the latter structure function is due to light 
partons ($u$,$d$,$s$ and $g$) only. Below $x = 2\times 10^{-3}$ the charm as 
well as the light parton contribution can become negative so that in this 
region we have taken the absolute values of $g_1$. Further we observe that 
the charm component of the total structure 
function $g_1^{\rm light}+g_1^{\rm approx}$ 
is small. At $Q^2=10$ $({\rm GeV}/c)^2$ (fig. 9a) it becomes maximal
at $x=10^{-3}$ where it amounts to $14 \%$ of the light parton contribution.
At larger $Q^2$-values i.e. $Q^2=50,100$ $({\rm GeV}/c)^2$ (see figs. 9b,9c)
the maximum occurs at $x=0.01$ where it is only $4 \%$. From these
results we conclude that the charm component of the structure function $g_1$
is much smaller than the one discovered for $F_2$ in \cite{lrsn1}. In the
latter case it becomes as large as $40 \%$ in the small $x$-region 
($x=10^{-4}$).
The above predictions made for $g_1(x,Q^2,m^2)$ up to NLO can be
considered as reliable because the bulk of the corrections are in the 
region $x<0.1$ at $Q^2$-values  for which $F_2^{\rm approx}$ and 
$g_1^{\rm approx}$ (Born) are close to their exact values. This region is
accessible to fixed target and HERA experiments where in the latter case both
the electron beam and proton beam are polarized.

Summarizing our work we have computed the order $\alpha_s^2$ contributions
to the heavy quark coefficient functions corresponding to the spin structure
function $g_1(x,Q^2,m^2)$. 
For the Compton process (2.17) we were able to calculate the
exact coefficient function $L_q^{\rm NS,(2)}$. In the case of the photon-gluon
fusion process (2.16) and the Bethe-Heitler process (2.17) we could only
obtain an analytic expression of $H_l^{(2)}$ ($l=q,g$) in the
asymptotic regime $Q^2 \gg m^2$. The last expressions will serve as a check
on the exact coefficient functions which can be computed in a semi-analytic way
similar to the procedure outlined in \cite{lrsn1}. To estimate the NLO 
corrections to $g_1(x,Q^2,m^2)$ we modified the asymptotic coefficient functions
according to (5.1), (5.3) in order to mimic the exact form. This approximation
was tested for $F_2$ and $g_1$ (Born) and lead to reasonable results as long 
as $Q^2>10$ $({\rm GeV}/c)^2$ and $x<0.1$. Using this approximation we
found that the order $\alpha_s$ corections to $g_1(x,Q^2,m^2)$ are large but 
the charm component of the total spin structure function given by 
$g_1^{\rm light}+g_1^{\rm approx}$ is small.

\appendix
\mysection*{Appendix A}
\setcounter{section}{1}

Here we present the unrenormalized spin dependent operator matrix elements 
$\hat A_{ij}^{(2)}$ whose general structure was expressed in 
renormalization group coefficients in section 3. 
After having carried out mass
renormalization the two-loop OME is given by the
following expression (see also (3.22)) 
\begin{eqnarray}
&&\hat A_{Qg}^{(2)}\Big(\frac{m^2}{\mu^2},\epsilon\Big)
=S_{\epsilon}^2\Big(\frac{m^2}
{\mu^2}\Big)^{\epsilon}
\Big[\frac{1}{\epsilon^2}\Big\{C_FT_f\Big[(16-32z)[\ln z-2\ln(1-z)]
+24\Big]
\nonumber \\ && \qquad
+C_AT_f[(32 - 64 z )\ln(1-z)
-64(1 + z)\ln z-192(1-z)
]\Big\}
\nonumber \\ && \qquad
+\frac{1}{\epsilon}\Big\{C_FT_f \Big[(4-8z)[
2\ln^2(1-z)+\ln^2z-4\ln z\ln(1-z)-4\zeta(2)]
\nonumber \\ && \qquad
-32(1-z)\ln(1-z)
+(4-64z)\ln z  - 8-12z\Big]
\nonumber \\ && \qquad
+ C_AT_f\Big[(8+16z)[2{\rm Li}_2(-z) + 2\ln z\ln(1+z)+\ln^2z]
\nonumber \\ && \qquad
-(8 - 16 z)\ln^2(1-z)
+16\zeta(2)+32(1-z)\ln(1-z)
\nonumber \\ && \qquad
-(8+64z)\ln z-96+88z \Big]\Big\}
+a^{(2)}_{Qg}(z)\Big]
\nonumber \\ && \qquad
+\sum_{H=Q}^tS_{\epsilon}^2\Big(
\frac{m_H^2}{\mu^2}\Big)^{\epsilon/2} \Big( \frac{m^2}{\mu^2}\Big)^{\epsilon/2}
\Big[\frac{1}{\epsilon^2}T_f^2\Big(
\frac{64}{3}-\frac{128}{3}z\Big)
\nonumber \\ && \qquad \qquad
\times \Big(1 + \frac{\epsilon^2}{4}\zeta(2)\Big)\Big]
 \,,
\end{eqnarray}
with 
\begin{eqnarray}
&&a^{(2)}_{Qg}(z)=C_FT_f\Big\{(-1+2z)[8\zeta(3)+8\zeta(2)\ln(1-z)
+\frac{4}{3}\ln^3(1-z)
\nonumber \\ && \qquad
-8\ln(1-z){\rm Li}_2(1-z)
+4\zeta(2)\ln z
-4\ln z\ln^2(1-z)
\nonumber \\ && \qquad
+\frac{2}{3}\ln^3z
-8\ln z{\rm Li}_2(1-z)
+8{\rm Li}_3(1-z)
-24{\rm S}_{1,2}(1-z)]
\nonumber \\ && \qquad
-(116-48z-16z^2){\rm Li}_2(1-z)
+(50-32z-8z^2)\zeta(2)
\nonumber \\ && \qquad
-(72-16z-8z^2)\ln z\ln(1-z)
+(12-8z-4z^2)\ln^2(1-z)
\nonumber \\ && \qquad
-(5-8z-4z^2)\ln^2z-(64-60z)\ln(1-z)
\nonumber \\ && \qquad
-(16+50z)\ln z
-22+46z\Big\}
\nonumber \\ && \qquad
+C_AT_f\Big\{(-1+2z) [-8\zeta(2)\ln(1-z)
-\frac{4}{3} \ln^3(1-z)
\nonumber \\ && \qquad
+8\ln(1-z){\rm Li}_2(1-z)-8{\rm Li}_3(1-z)]
+(1+2z)\Big[-\frac{4}{3}\ln^3z
\nonumber \\ && \qquad
-8\zeta(2)\ln(1+z)-16\ln(1+z){\rm Li}_2(-z)
-8\ln z\ln^2(1+z)
\nonumber \\ && \qquad
+4\ln^2z\ln(1+z)+ 8\ln z{\rm Li}_2(-z)-8{\rm Li}_3(-z)
-16{\rm S}_{1,2}(-z)\Big]
\nonumber \\ && \qquad
+16(1+z)[4{\rm S}_{1,2}(1-z)
+2\ln z{\rm Li}_2(1-z)-3\zeta(2)\ln z+{\rm Li}_2(-z)
\nonumber \\ && \qquad
+\ln z\ln(1+z)]
-16(1-z)\zeta(3)
+(100-112z-8z^2)
{\rm Li}_2(1-z)
\nonumber \\ && \qquad
-(132-144z-4z^2)\zeta(2)
-4z(4+z)\ln z\ln(1-z)
\nonumber \\ && \qquad
-(10-8z-2z^2)\ln^2(1-z)-(6+2z^2)\ln^2z
\nonumber \\ && \qquad
+4\ln(1-z)-(56+148z)\ln z
-204+212z\Big\} \,.
\end{eqnarray}

The unrenormalized OME corresponding to fig. 3 (see (3.28))
is given by
\begin{eqnarray}
&&\hat A^{{\rm PS},(2)}_{Qq}\Big(\frac{m^2}{\mu^2},\epsilon\Big)=S_{\epsilon}^2
\Big(\frac{m^2}{\mu^2}\Big)^{\epsilon}C_FT_f\Big\{
\frac{1}{\epsilon^2}\Big(-32(1+z)\ln z-80(1-z)\Big)
\nonumber \\ &&  
+\frac{1}{\epsilon}\Big(8(1+z)\ln^2z+8(1-3z)\ln z
-8(1-z)\Big)
+a^{{\rm PS},(2)}_{Qq}(z)\Big\} \,,
\end{eqnarray}
with
\begin{eqnarray}
&&a^{{\rm PS},(2)}_{Qq}(z)=
 (1+z)[32{\rm S}_{1,2}(1-z)+16\ln z{\rm Li}_2(1-z)
-24\zeta(2)\ln z
\nonumber \\ && \qquad
-\frac{4}{3}\ln^3z]
+20(1-z)[2{\rm Li}_2(1-z)-3\zeta(2)]
-(2-6z)  \ln^2z
\nonumber \\ && \qquad
-(12+60z)\ln z
-72(1-z)  \,.
\end{eqnarray}


\mysection*{Appendix B}
\setcounter{section}{2}

In this appendix we present the spin dependent heavy quark coefficient 
functions $H_{i}^{(2)}$ and $L_{i}^{(2)}$ ($i=q,g$) in the asymptotic
limit $Q^2\gg m^2$. Starting with the lowest order photon-gluon fusion process
$(2.13)$ the heavy quark coefficient function reads (see (4.3))

\begin{eqnarray}
 && H^{(1)}_{g}\Big(z,\frac{Q^2}{m^2},\frac{m^2}{\mu^2}\Big)=
T_f[- 4 (1 - 2 z ) \Big( \ln\frac{Q^2}{m^2}
+\ln(1-z) - \ln z\Big)
\nonumber \\ && \qquad
+ 4 (3 - 4 z)] \,.
\end{eqnarray}

In next-to-leading order the coefficient function corresponding
to the virtual corrections to the Born reaction $(2.13)$ and the
bremstrahlung process $(2.16)$ are given by (see (4.6))
\begin{eqnarray}
 && H^{(2)}_{g}\Big(z,\frac{Q^2}{m^2},\frac{m^2}{\mu^2}\Big)=
 \Big[C_F T_f \{  (8 z - 4) (2 \ln(1 - z) - \ln z) + 6 \}
\nonumber \\ && \qquad
+ C_A T_f \{ - 8 (1 - 2 z)  \ln(1 - z) + 16 (1 + z) \ln z + 48 (1 - z)\}\Big]
\ln^2 \frac{Q^2}{m^2}
\nonumber \\ && \qquad
+\Big[ C_F T_f \{ -8 (1 - 2 z) [{\rm Li}_2(1-z) - 3 \ln z \ln(1 - z)
\nonumber \\  && \qquad
+ 2 \ln^2(1 - z) + \ln^2 z - 4 \zeta(2)] 
\nonumber \\ && \qquad
+ 4 (17 - 20 z) \ln(1 - z) - 16 (3 - 2 z) \ln z - 4 (17 - 13 z) \}
\nonumber \\ && \qquad
+ C_A T_f \{ - 16 (1 + 2 z) [{\rm Li}_2(-z) + \ln z \ln(1 + z)]
\nonumber \\ && \qquad
+ 32 (1 + z) {\rm Li}_2(1-z) + 48 \ln z \ln( 1- z) - 8(1 - 2 z) \ln^2(1 - z)
\nonumber \\ && \qquad
- 8(3 + 4 z) \ln^2 z - 32 z \zeta(2) + 16 (7 - 8 z) \ln(1 - z)
\nonumber \\ && \qquad
- 24 (5 - 4 z) \ln z - 8 (20 - 21 z) \} \Big] \ln\frac{Q^2}{m^2}
\nonumber \\ && \qquad
+ C_A T_f \Big[\{-16(1 - 2 z) \ln(1 - z) + 32 (1 + z) \ln z 
+ 96 (1 - z)\} \ln \frac{Q^2}{m^2}
\nonumber \\ && \qquad
+ 32 (1 + z) {\rm Li}_2(1-z) + 48 \ln z \ln(1 - z) 
- 16 (1 - 2 z) \ln^2(1 - z) 
\nonumber \\ && \qquad
- 16 (1 + z) \ln^2 z + 16 (9 - 10 z) \ln(1 - z) 
\nonumber \\ && \qquad
- 32 (4 - z) \ln z
+ 16 (1 - 2 z) \zeta(2) - 256 (1 - z) \Big] \ln \frac{m^2}{\mu^2}
\nonumber \\ && \qquad
+ C_F T_f \Big[ (1 - 2 z) [24 {\rm Li}_3(1-z) - 8 \ln(1 - z) {\rm Li}_2(1-z)
\nonumber \\ && \qquad
- 32 \zeta(2) \ln z - 8 \ln^3(1 - z) + 20 \ln z \ln^2(1 - z)
\nonumber \\ && \qquad
- 16 \ln^2 z \ln(1 - z) + \frac{8}{3} \ln^3 z ]
\nonumber \\ && \qquad
- 16 (1 + z)^2 [4 {\rm S}_{1,2}(-z) + 4 \ln(1 + z) {\rm Li}_2(-z)
\nonumber \\ && \qquad
+ 2 \ln z \ln^2(1 + z) - \ln^2 z \ln(1 + z) + 2 \zeta(2) \ln(1 + z)]
\nonumber \\ && \qquad
- (32 - 192  z + 32  z^2) {\rm Li}_3(-z) - (96 - 16  z^2) {\rm Li}_2(1-z)
\nonumber \\ && \qquad
+ 32  (1 - z)^2  [{\rm S}_{1,2}(1-z) + \ln z {\rm Li}_2(-z)]
\nonumber \\ && \qquad
+ (\frac{64}{3 z} + 64 z  + \frac{208}{3} z^2) 
[{\rm Li}_2(-z) + \ln z \ln(1 + z)]
\nonumber \\ && \qquad
+ (66 - 80  z - 4 z^2) \ln^2(1 - z) - 32 z^2 \zeta(2) \ln(1 - z)
\nonumber \\ && \qquad
- (188 - 164 z - 16 z^2) \ln(1 - z) 
+ (36 - 8 z - \frac{92}{3} z^2) \ln^2 z
\nonumber \\ && \qquad
- (160 - 112 z - 8 z^2) \ln z \ln(1 - z)
+ \frac{1}{3} (320 - 424 z - 48 z^2) \ln z
\nonumber \\ && \qquad
- (48 - 224 z - 32 z^2) \zeta(3) - \frac{1}{3}(192 - 336 z - 184 z^2) \zeta(2)
\nonumber \\ && \qquad
+ \frac{1}{3}(304 - 244 z) \Big]
\nonumber \\ && \qquad
+ C_A T_f \Big[ 16 (1 + 2 z) \Big({\rm Li}_3 \Big( \frac{1-z}{1+z}\Big)
- {\rm Li}_3\Big( - \frac{1-z}{1+z}\Big)
\nonumber \\ && \qquad
- \ln(1 - z) {\rm Li}_2(-z) - \ln z \ln(1 - z) \ln(1 + z) \Big)
\nonumber \\ && \qquad
+ 8 (1 + 2 z + 2 z^2) [2 {\rm S}_{1,2}(-z) + \ln z \ln^2(1 + z)
\nonumber \\ && \qquad
+ 2 \ln(1 + z) {\rm Li}_2(-z) + \zeta(2) \ln(1 + z)]
\nonumber \\ && \qquad
+ (72 + 80 z - 16 z^2) {\rm S}_{1,2}(1-z) - 16 (2 + z) [2 {\rm Li}_3(1-z)
\nonumber \\ && \qquad
+ \ln^2 z \ln(1 - z)] - (12 + 24 z - 8 z^2) [2 {\rm Li}_3(-z)
\nonumber \\ && \qquad
- \ln^2 z \ln(1 + z) - 2 \ln z {\rm Li}_2(-z)]
\nonumber \\ && \qquad
- \frac{16}{3} \Big(\frac{2}{z} + 3 + 9 z + 11 z^2\Big) 
[{\rm Li}_2(-z) + \ln z \ln(1 + z)]
\nonumber \\ && \qquad
+ 24 \ln z \ln^2(1 - z) + 16 \ln z {\rm Li}_2(1-z) 
\nonumber \\ && \qquad
+ 32 (1 + z) \ln(1 - z) {\rm Li}_2(1-z) + (76 - 112 z - 8 z^2) {\rm Li}_2(1-z)
\nonumber \\ && \qquad
+ (38 - 48 z + 2 z^2) \ln^2(1 - z) + (24 - 80 z + 16 z^2) \zeta(2) \ln(1 - z)
\nonumber \\ && \qquad
+ \frac{8}{3}(3 + 4 z) \ln^3 z + \Big(84 - 32 z + \frac{82}{3} z^2\Big)\ln^2 z
\nonumber \\ && \qquad
- (136 - 112 z + 4 z^2) \ln z \ln(1 - z) - (80 + 32 z) \zeta(2) \ln z
\nonumber \\ && \qquad
+ \frac{1}{3} (776 - 652 z + 24 z^2) \ln z
- (172 - 212 z + 8 z^2) \ln(1 - z)
\nonumber \\ && \qquad
- (28 + 24 z + 16 z^2) \zeta(3) 
- \Big(228 - 224 z + \frac{164}{3} z^2 \Big) \zeta(2)
\nonumber \\ && \qquad
+ \frac{1}{3} (808 - 832 z) \Big]\,.
\end{eqnarray}

The coefficient function corresponding to the Bethe-Heitler process
$(2.17)$ reads (see (4.9))

\begin{eqnarray}
 && H^{(2)}_{q}\Big(z,\frac{Q^2}{m^2},\frac{m^2}{\mu^2}\Big)=
C_F T_f\Big[ \Big\{ 8 (1+z) \ln z 
+ 20 (1 - z) \Big\}
\ln^2 \frac{Q^2}{m^2}
\nonumber \\ && \qquad
+ \Big\{[ 16 (1+z) \ln z + 40 (1 - z) ]
\ln \frac{Q^2}{m^2}
\nonumber \\ && \qquad
+ 8 ( 1+z)[ 2 {\rm Li}_2(1-z) + 2 \ln z \ln (1-z) - \ln^2 z] + 40 (1 - z)
\nonumber \\ && \qquad
\times \ln(1-z)
- (56 - 8 z) \ln z  - 96 (1 - z)\Big\}
\ln\frac{m^2}{\mu^2}
\nonumber \\ && \qquad
+ \{ 16 ( 1+z) [ {\rm Li}_2(1-z) + \ln z \ln (1-z) - \ln^2 z]
\nonumber \\ && \qquad
+ 40 ( 1- z) \ln(1-z)
- 32 (2 - z) \ln z - 88 (1 - z)
\} \ln \frac{Q^2}{m^2}
\nonumber \\ && \qquad
+(1+z) \Big( 32 \,{\rm S}_{1,2}(1-z) - 16 {\rm Li}_3(1-z) + 8 \ln z \ln^2(1-z)
\nonumber \\ && \qquad
-16 \ln^2 z \ln(1-z) + 16 \ln(1-z) {\rm Li}_2(1-z)
- 32 \zeta(2) \ln z 
\nonumber \\ && \qquad
+\frac{16}{3}\ln^3 z\Big) + 16 (1 - 3 z) {\rm Li}_2(1-z)
\nonumber \\ && \qquad
-\Big(\frac{32}{3z} + 32 + 32 z + \frac{32}{3}z^2\Big) [ {\rm Li}_2(-z)
+ \ln z \ln(1+z)]
\nonumber \\ && \qquad
-\Big(112 - 80 z + \frac{32}{3}z^2\Big) \zeta(2)
- 32 (2 - z) \ln z \ln (1-z)
\nonumber \\ && \qquad
+ (1 - z) \Big(20 \ln^2(1 - z) - 88 \ln(1 - z) 
+ \frac{592}{3}\Big)
\nonumber \\ && \qquad
+ \Big(56 + \frac{16}{3} z^2\Big) \ln^2 z 
+ \frac{256}{3} \Big(2 -  z \Big) \ln z 
 \Big]   \,.
\end{eqnarray}

Finally we present the coefficient function originating from the
Compton process $(2.17)$. The asymptotic form is given by $(4.12)$
and can be analytically expressed as

\begin{eqnarray}
 && L^{{\rm NS},(2)}_{q}\Big(z,\frac{Q^2}{m^2},\frac{m^2}{\mu^2}\Big)=
C_F T_f \Big[  \frac{4}{3} \left ( \frac{1+z^2}{1-z} \right )
 \ln^2 \frac{Q^2}{m^2}
+ \Big\{ \frac{1+z^2}{1-z} \Big( \frac{8}{3} \ln (1-z)
\nonumber \\ && \qquad
- \frac{16}{3} \ln z - \frac{58}{9} \Big)
- 2 + 6 z \Big\} \ln \frac{Q^2}{m^2}
\nonumber \\ && \qquad
+ \Big( \frac{1+z^2}{1-z} \Big) \Big ( - \frac{8}{3} {\rm Li}_2(1-z)
- \frac{8}{3} \zeta(2) - \frac{16}{3} \ln z \ln (1-z)
\nonumber \\ && \qquad
+ \frac{4}{3} \ln^2(1-z)
+ 4 \ln^2 z - \frac{58}{9} \ln(1-z) + \frac{134}{9} \ln z
+ \frac{359}{27} \Big )
\nonumber \\ && \qquad
-\Big(  2 - 6 z \Big) \ln(1-z)
+\Big(\frac{10}{3} - 10 z \Big) \ln z + \frac{19}{3} - 19 z \Big] \,.
\end{eqnarray}

In the above expression one should bear in mind that the singularity at $z=1$
will never show up because of the kinematical constraint
$z<Q^2/(Q^2 + 4 m^2)$. However after convoluting $L_q^{\rm NS,(2)}$ by
the parton densities, the structure function $g_1(x,Q^2,m^2)$ will
diverge as $\ln^3(Q^2/m^2)$ in the limit $Q^2\gg m^2$. In this limit the upper 
boundary $z_{\rm max}$ in $(2.9)$ will tend to one and the virtual gluon which
decays into the heavy quark pair becomes soft. The soft gluon annihilation
which causes the cubic logarithm above has to be added to the two-loop
vertex correction containing the heavy quark (Q) loop which is calculated
in appendix A of \cite{rn}. In this way the cubic logarithm is then cancelled.
The final result will be that in (B.4) the singular terms at $z=1$ have to be
replaced by the distributions $(\ln^k(1-z)/(1-z))_+$ defined by
\begin{eqnarray}
\int^1_0dz\ \Big(\frac{\ln^k(1-z)}{1-z}\Big)_+f(z)=\int_0^1dz\ \Big(
\frac{\ln^k(1-z)}{1-z}\Big)\{f(z)-f(1)\} \,,
\end{eqnarray}
and one has to add the following delta function contribution
to (B.4)
\begin{eqnarray}
&& L_q^{\rm NS,S+V,(2)}\Big(z,\frac{Q^2}{m^2},\frac{m^2}{\mu^2}\Big)=
C_FT_f\delta(1-z)\Big\{2\ln^2\Big(\frac{Q^2}{m^2}\Big)-\Big[
\frac{32}{3}\zeta(2)+\frac{38}{3}\Big]
\nonumber \\&& \qquad
\times \ln\Big(\frac{Q^2}{m^2}\Big)
+\frac{268}{9}\zeta(2)+\frac{265}{9}\Big\}\,.
\end{eqnarray}

%

\centerline{\bf \large{Figure Captions}}

\begin{description}
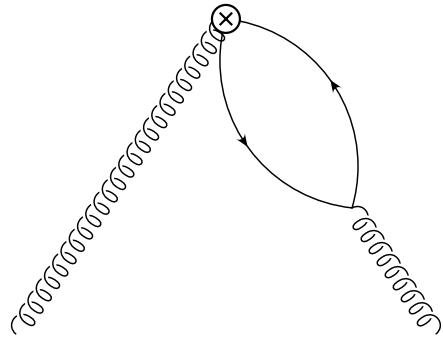
\item[Fig. 1.]
One-loop graphs contributing to the OME $A_{Qg}^{(1)}$.
The solid line indicates the heavy quark $Q$.\\
\item[Fig. 2.]
Two-loop graphs contributing to the OME $A_{Qg}^{(2)}$. The solid line
indicates the heavy quark $Q$. 
Graphs $d.19$ and $d.20$ contain the external gluon
self-energy with the heavy quark loop. In this loop a sum over all heavy
quark species indicated by H $(m_H^2 \ge m^2$) is understood.
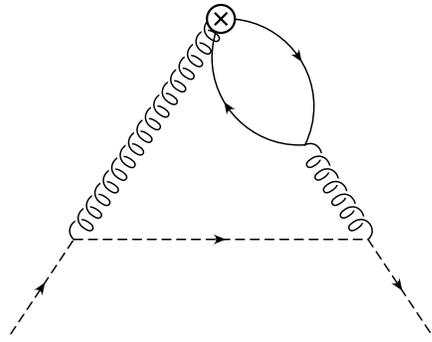
\item[Fig. 3.]
Two-loop graphs contributing to the OME $A_{Qq}^{{\rm PS},(2)}$. The solid line
represents the heavy quark $Q$ whereas the dashed line stands for the light
quark $q$.
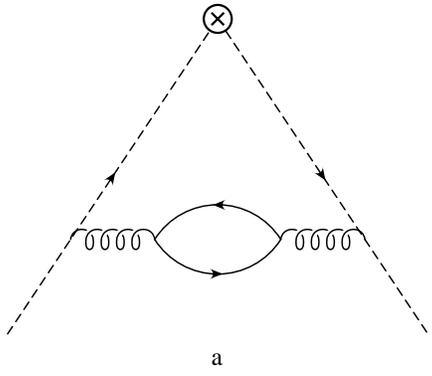
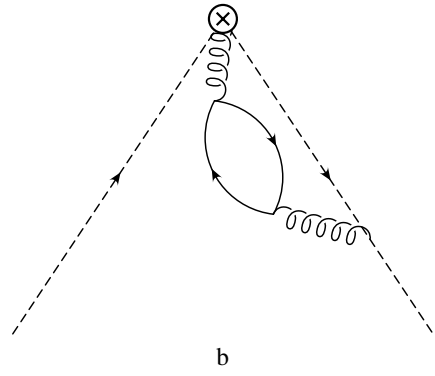
\item[Fig. 4.]
Two-loop graphs contributing to the OME $A_{qq,Q}^{{\rm NS},(2)}$.
The gluon self-energy contains
the heavy quark $Q$ with mass $m$ in the quark loop only
which is indicated by the solid line. The dashed line stands for the light
quark $q$.
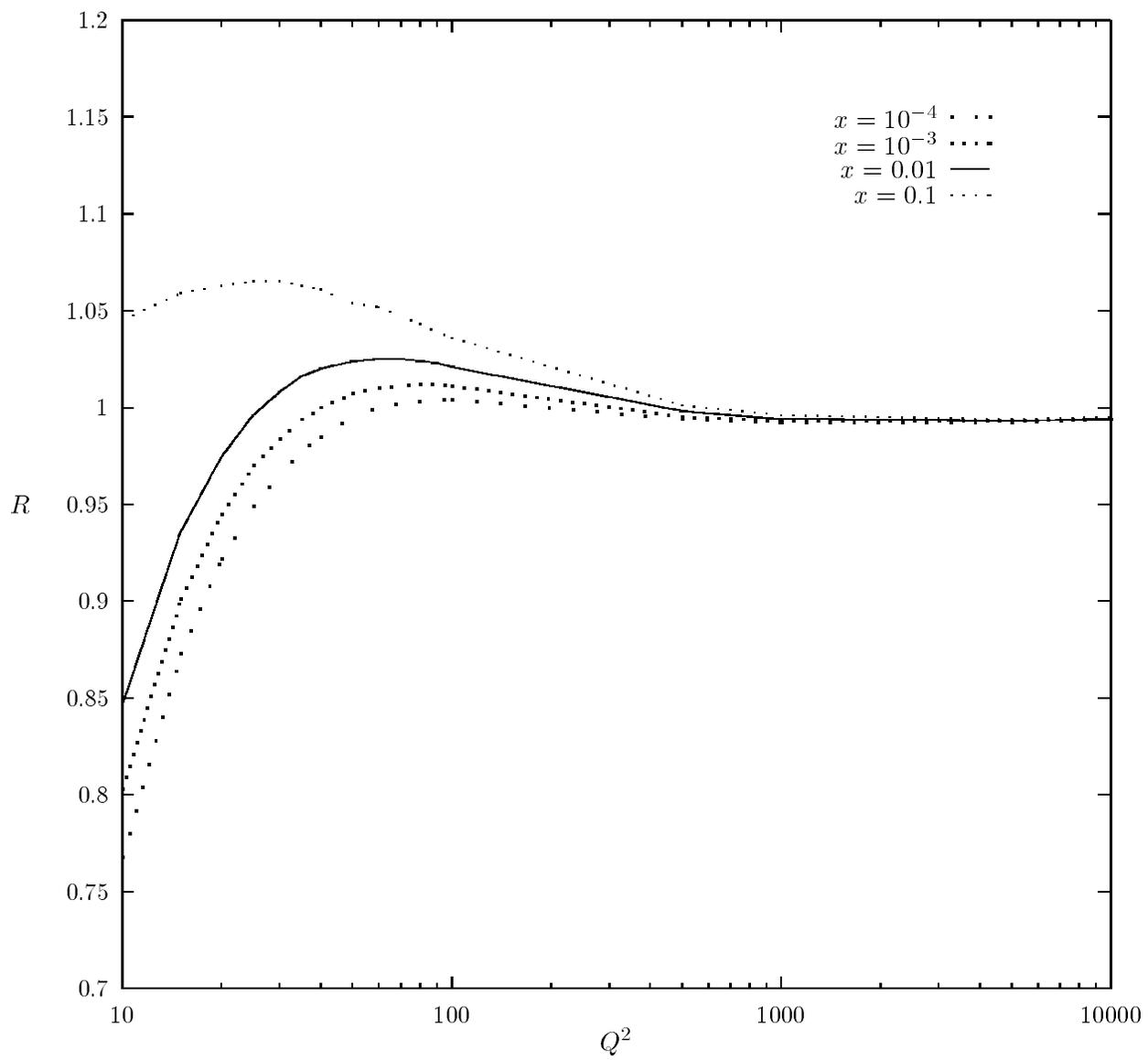
\item[Fig. 5.]
$R$ (5.5)
plotted as a function of $Q^2$ at fixed $x$;
$x = 0.1$  (upper dotted line), $x = 0.01$ (solid line), $x = 10^{-3}$ 
(middle dotted line),
$x = 10^{-4}$ (lower dotted line).
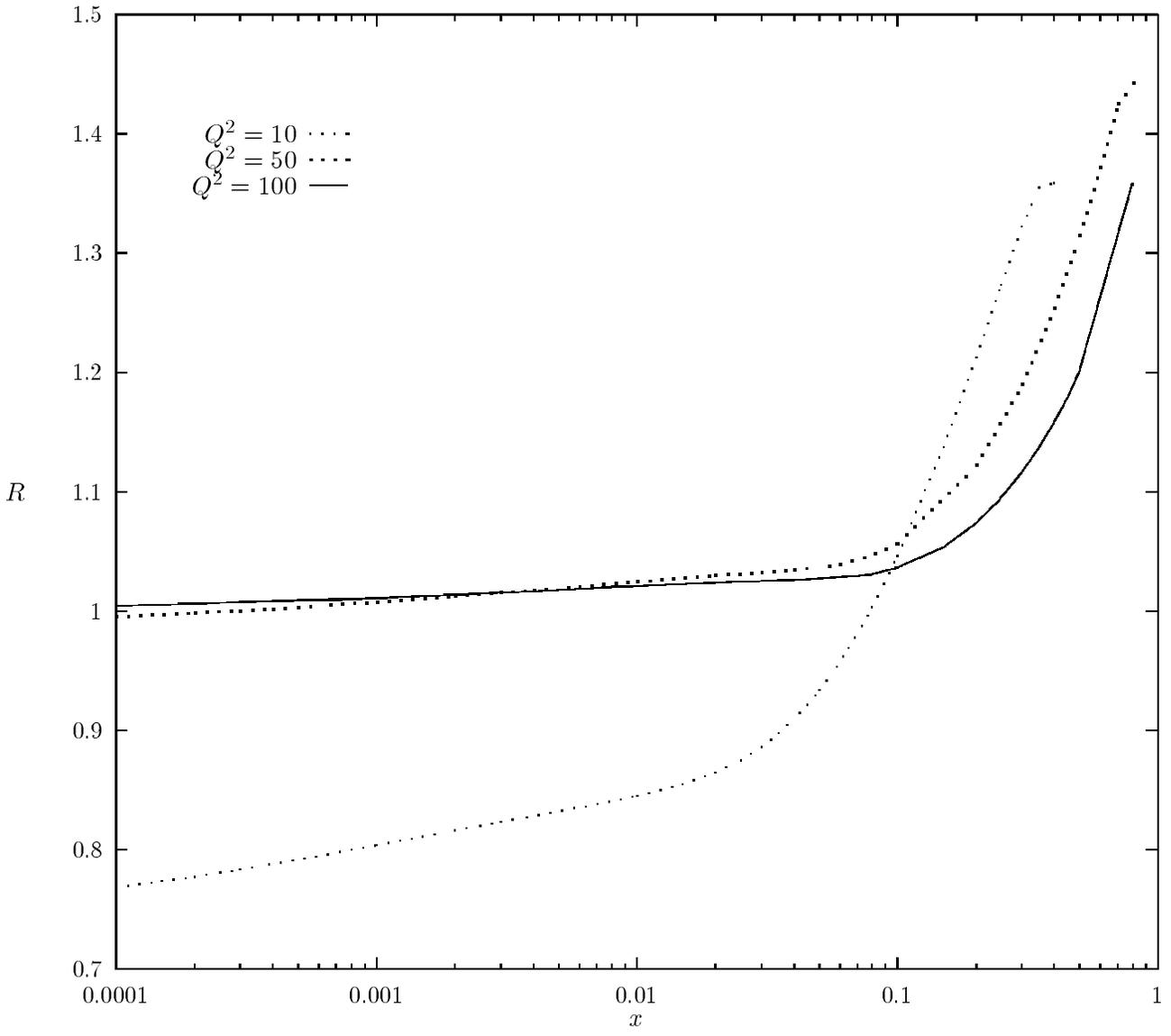
\item[Fig. 6.]
$R$ (5.5)
plotted as a function of $x$ at fixed $Q^2$;
$Q^2 = 10$ $({\rm GeV}/c)^2$ (lower dotted line), $Q^2 = 50$ $({\rm GeV}/c)^2$
(middle line), $Q^2 = 100$ $({\rm GeV}/c)^2$ (solid line).
\item[Fig. 7a.]
$g_1^{\rm exact}$ (Born) (dotted line) and $g_1^{\rm approx}$ (Born) (solid
line) as a function of $x$ for $Q^2 = 10$ $({\rm GeV}/c)^2$.
\item[Fig. 7b.]
Same as in Fig. 7a but now for $Q^2 = 50$ $({\rm GeV}/c)^2$.
\item[Fig. 7c.]
Same as in Fig. 7a but now for $Q^2 = 100$ $({\rm GeV}/c)^2$.
\item[Fig. 8a.]
$g_1^{\rm exact}$ (Born) (dotted line) and $g_1^{\rm approx}$ (NLO) (solid
line) as a function of $x$ for $Q^2 = 10$ $({\rm GeV}/c)^2$.
\item[Fig. 8b.]
Same as in Fig. 8a but now for $Q^2 = 50$ $({\rm GeV}/c)^2$.
\item[Fig. 8c.]
Same as in Fig. 8a but now for $Q^2 = 100$ $({\rm GeV}/c)^2$.
\item[Fig. 9a.]
$g_1^{\rm light}$ (solid line) and $g_1^{\rm approx}$ (dotted line)
both in NLO as a function of $x$ for $Q^2 = 10$ $({\rm GeV}/c)^2$.
For $x < 10^{-3}$ both $g_1^{\rm light}$ and $g_1^{\rm approx}$ become
negative so that we have taken their absolute values.
\item[Fig. 9b.]
Same as in Fig. 9a but now for $Q^2 = 50$ $({\rm GeV}/c)^2$.
\item[Fig. 9c.]
Same as in Fig. 9a but now for $Q^2 = 100$ $({\rm GeV}/c)^2$.
\end{description}

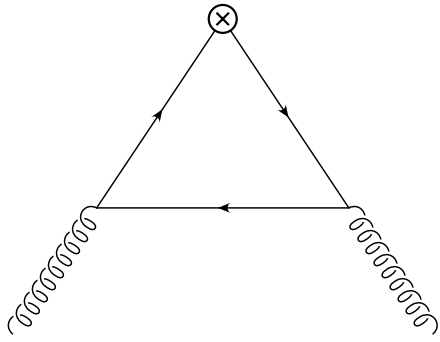
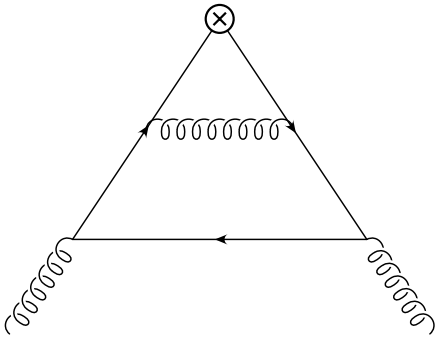
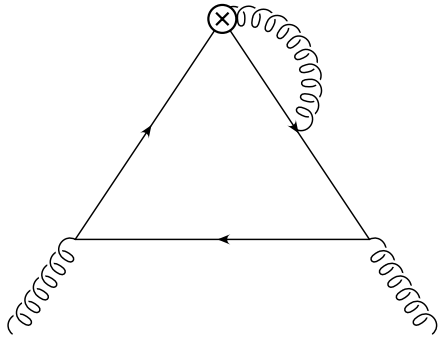
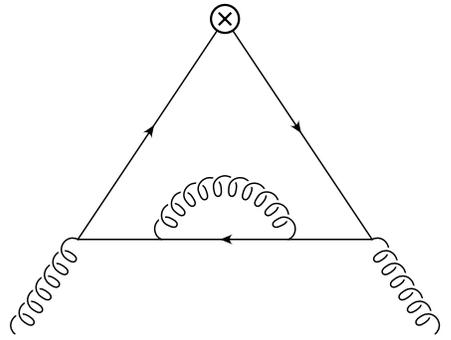
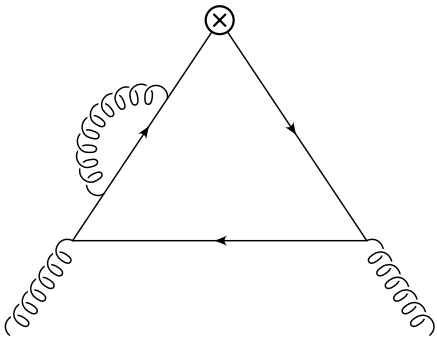
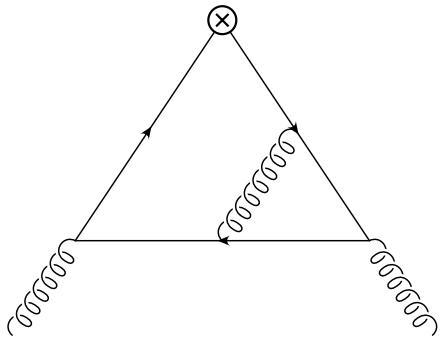
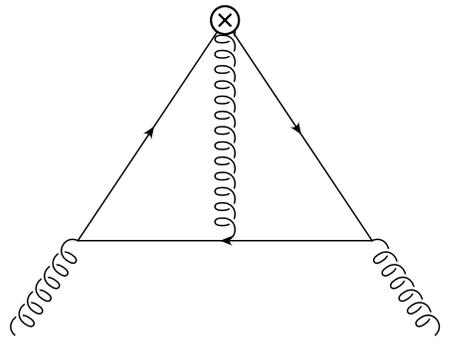
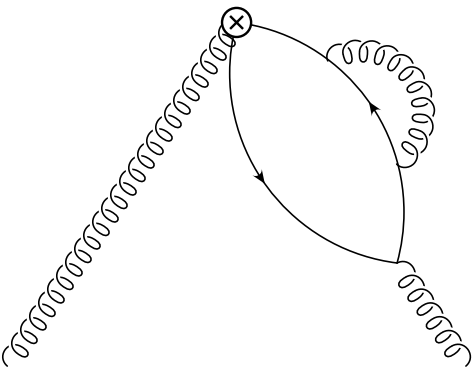
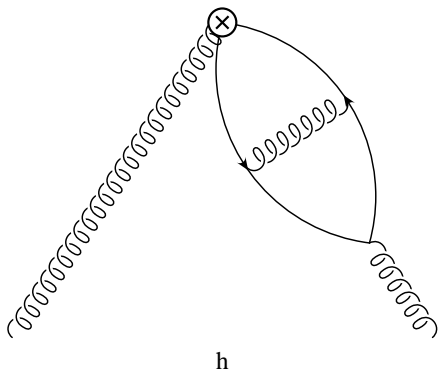
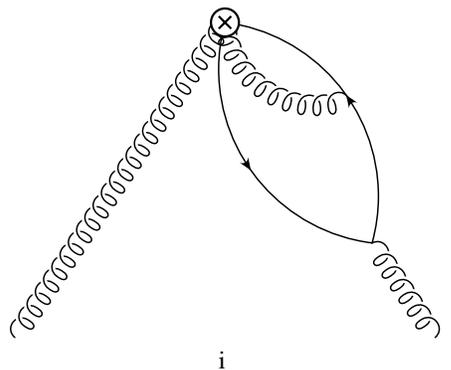
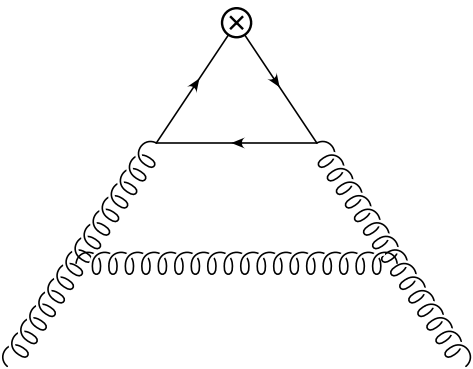
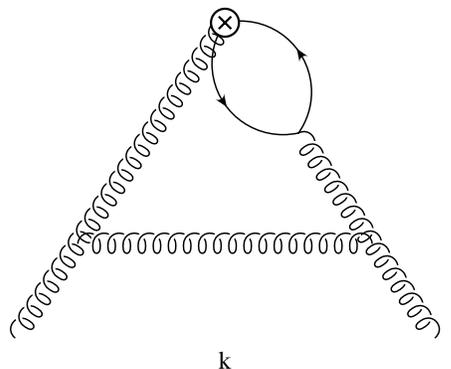
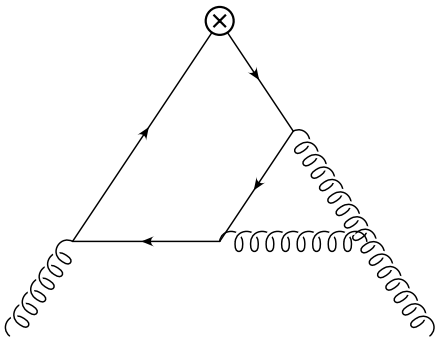
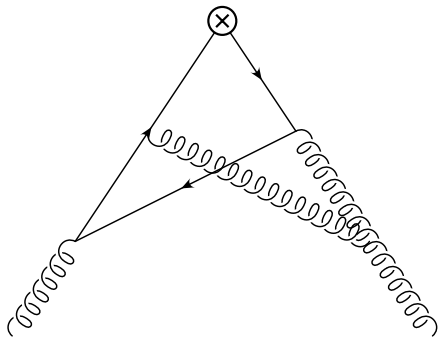
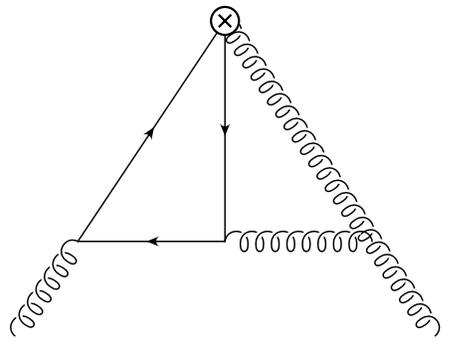
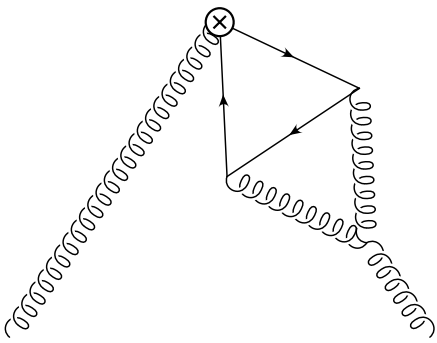
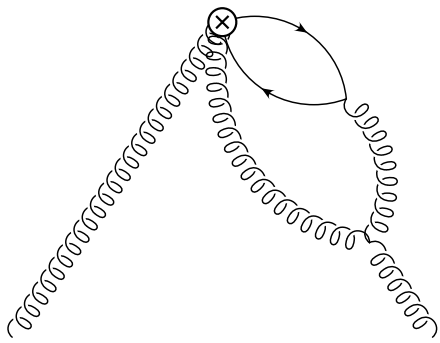
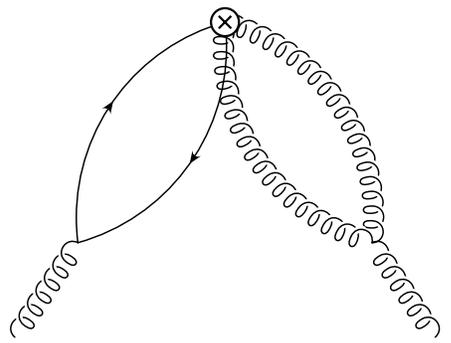
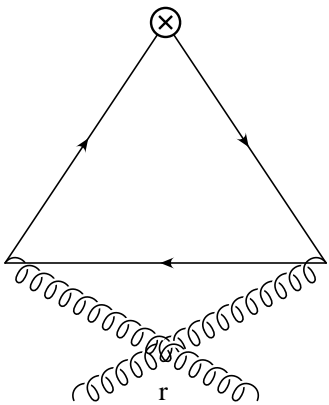
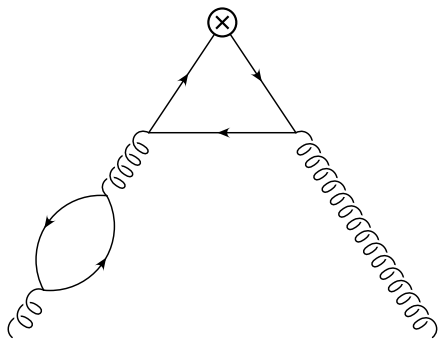
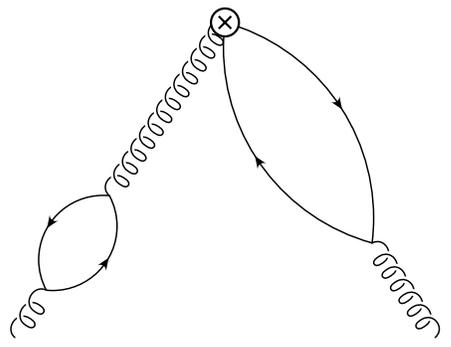
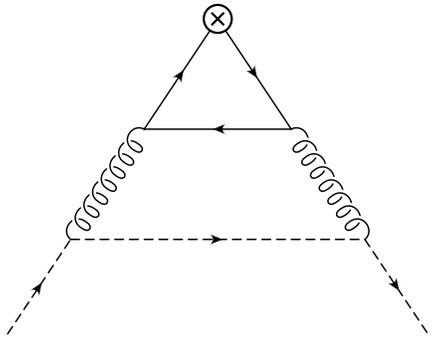
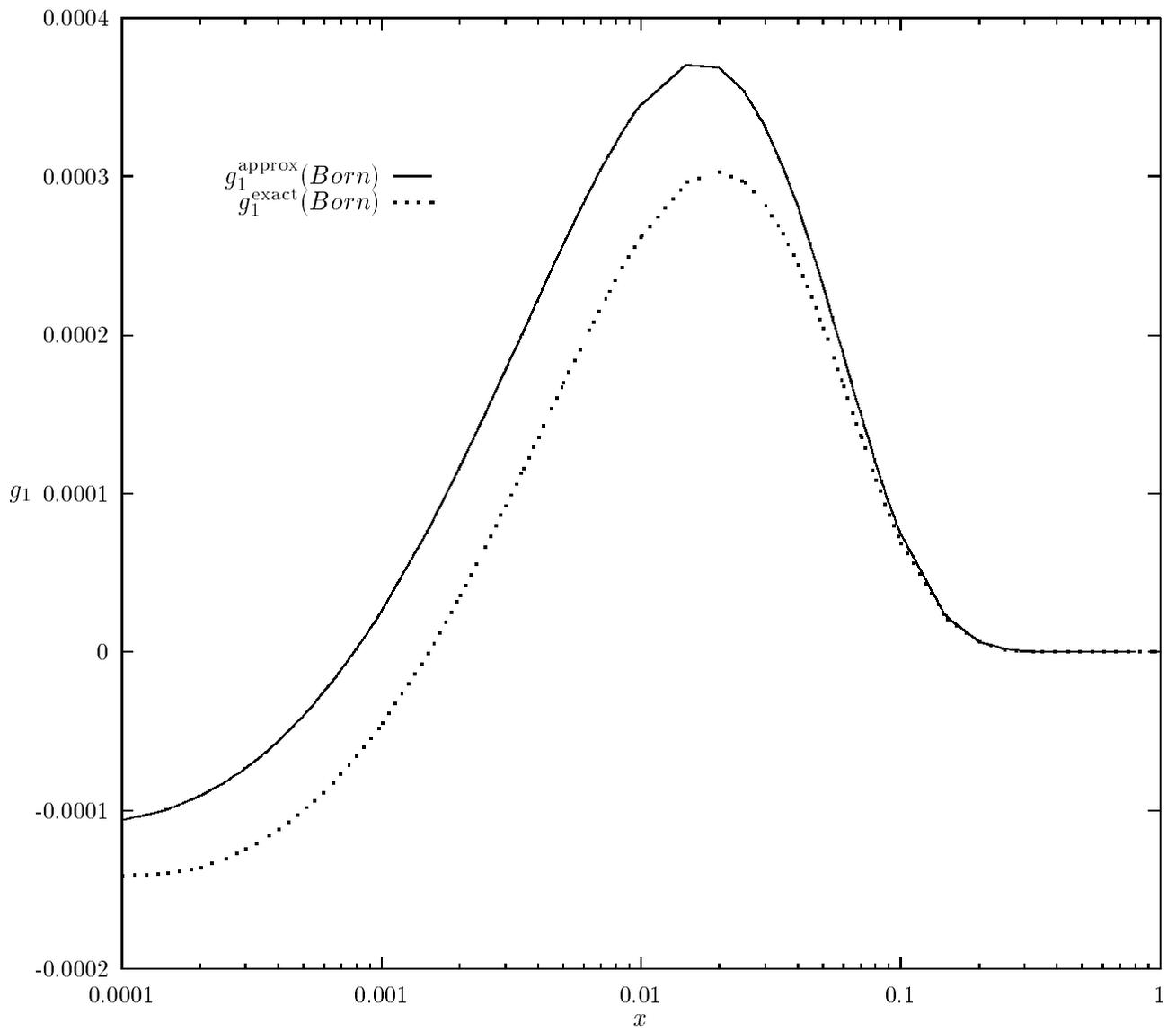
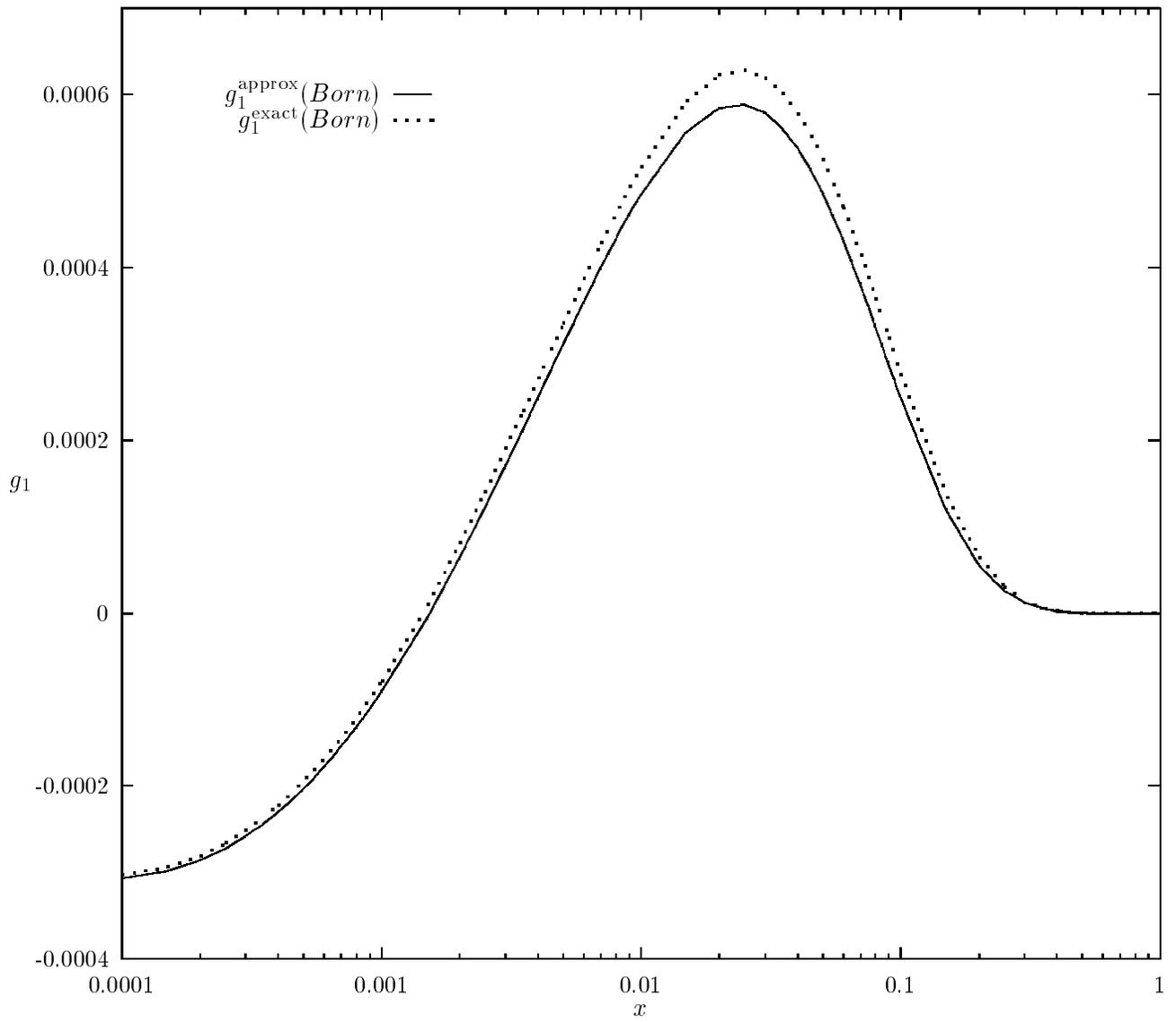
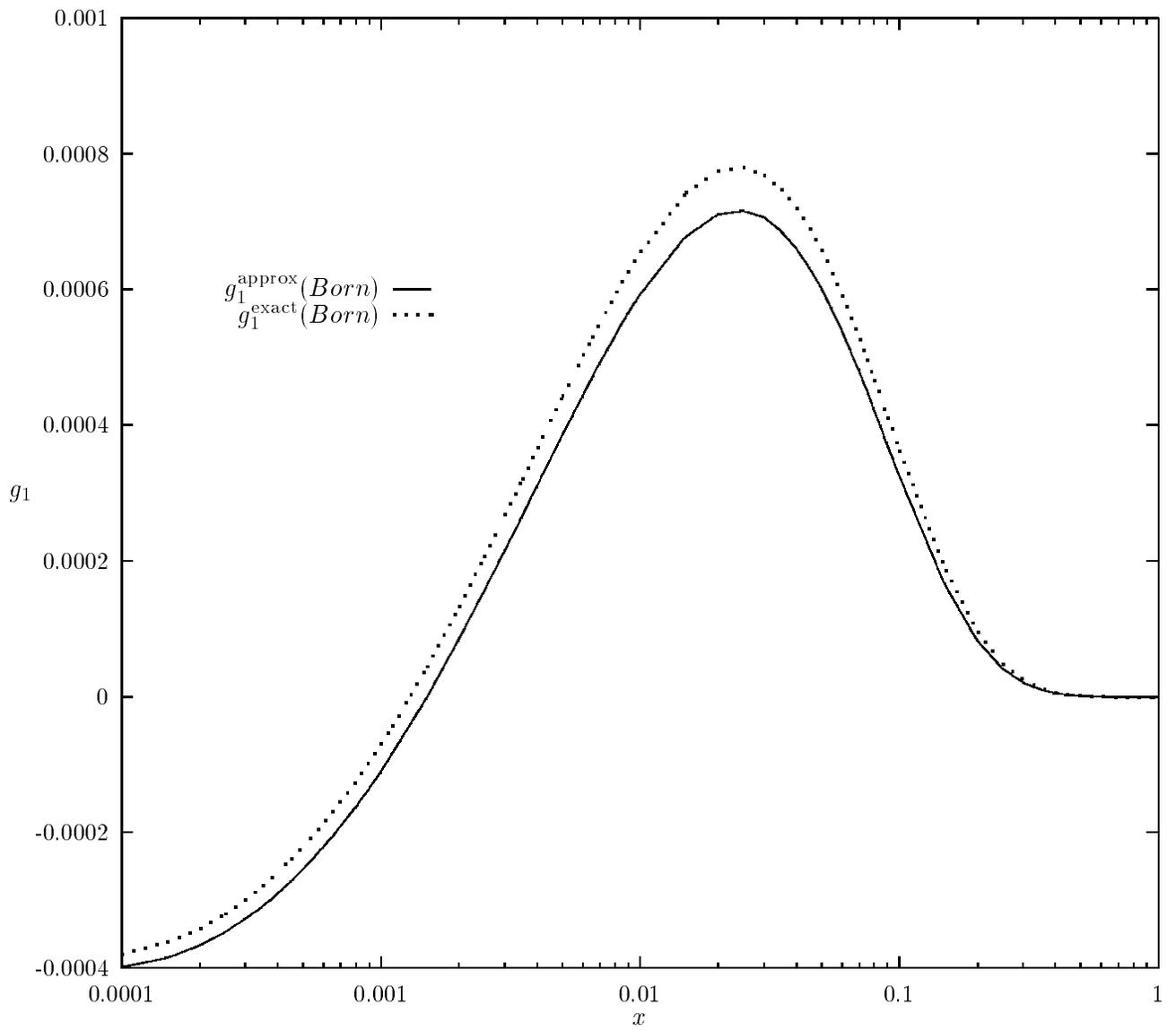
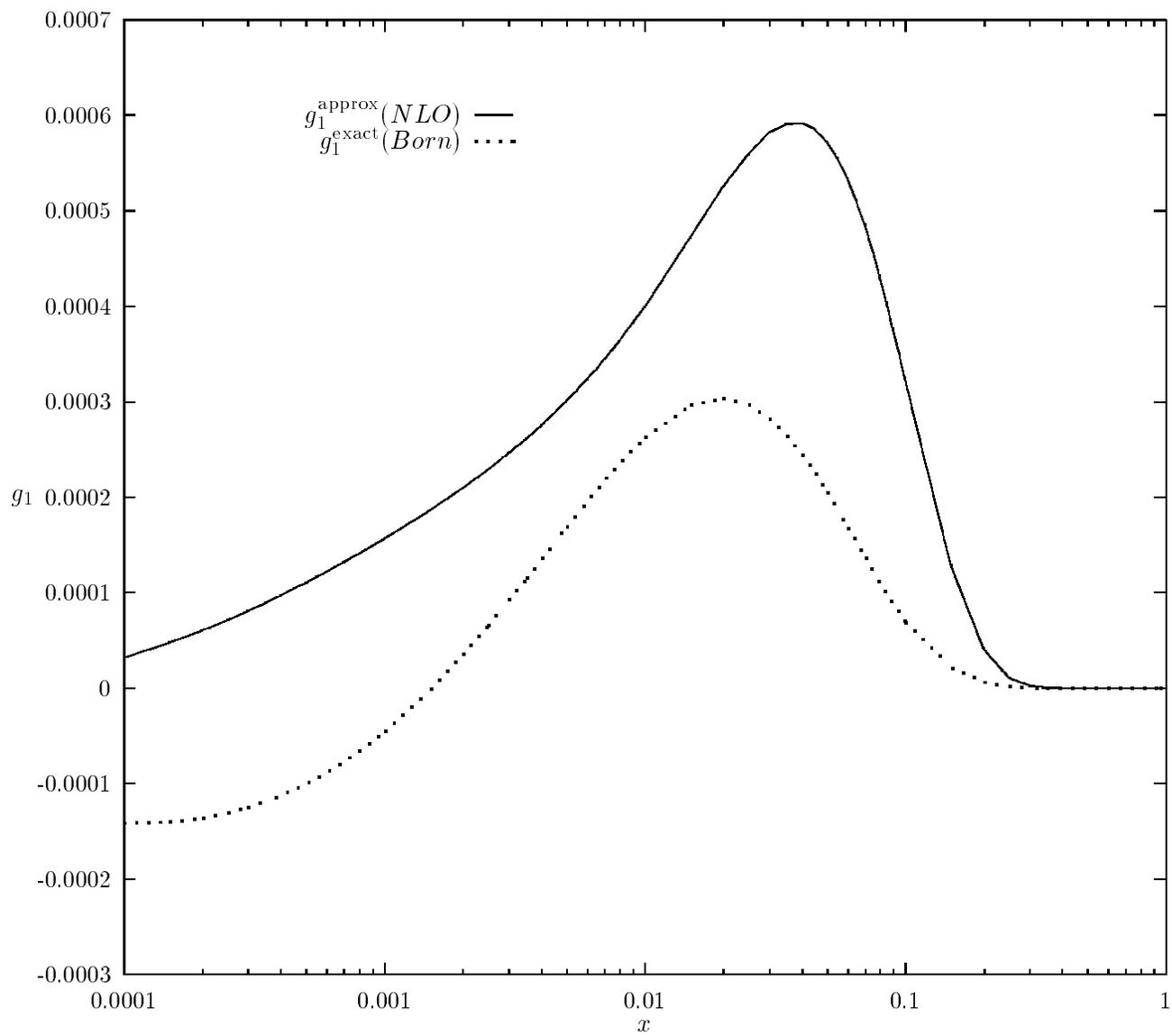
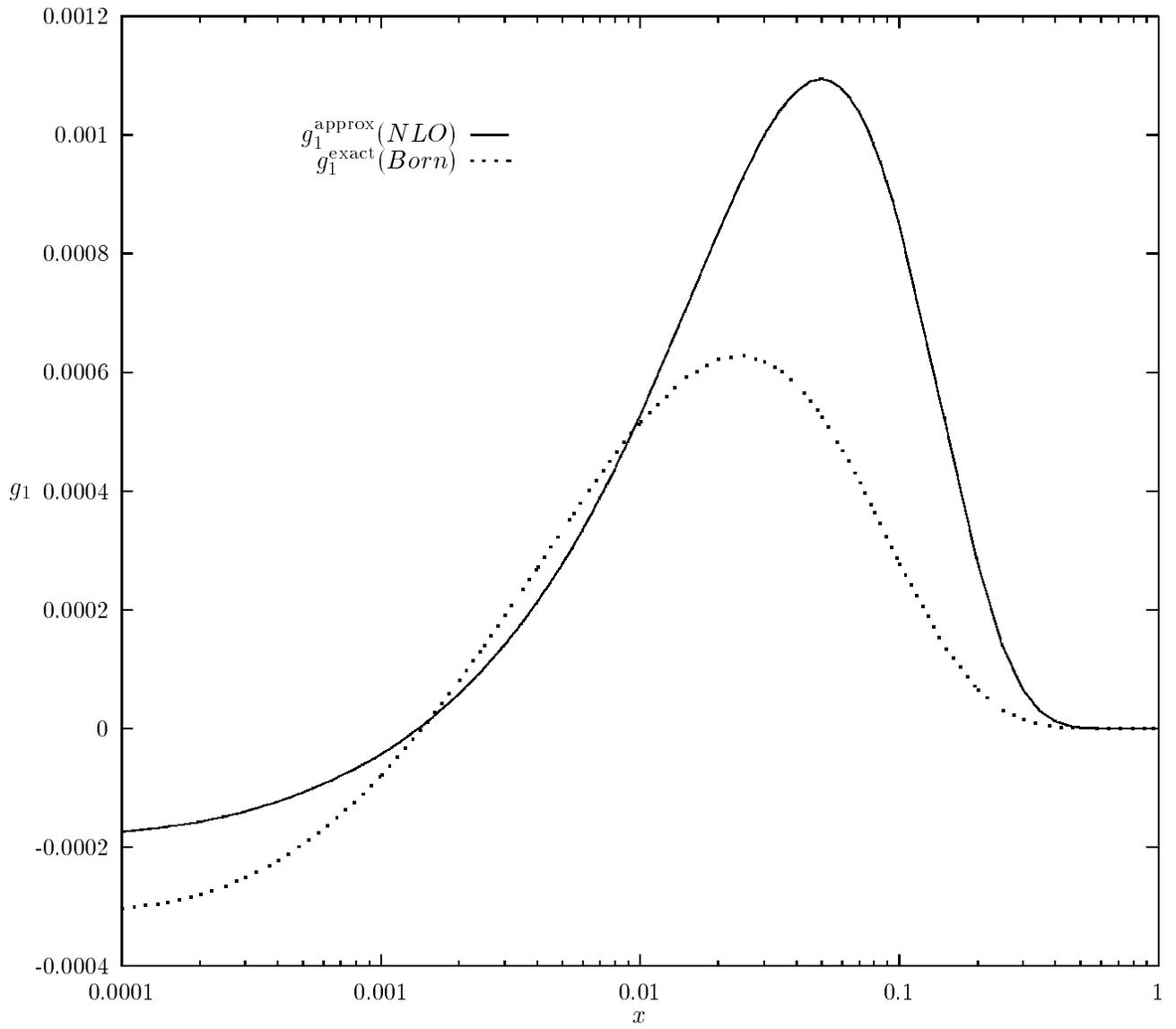
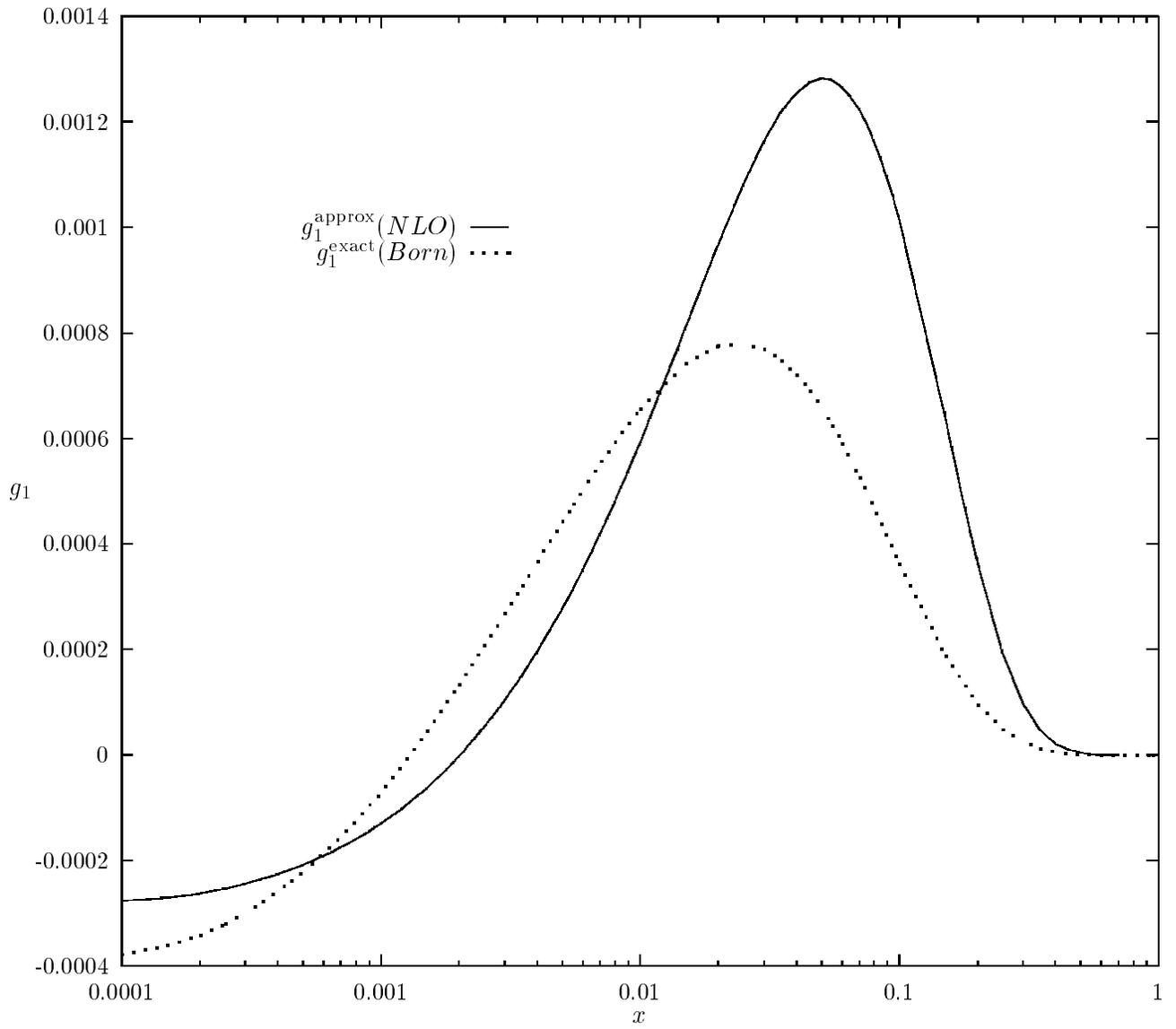
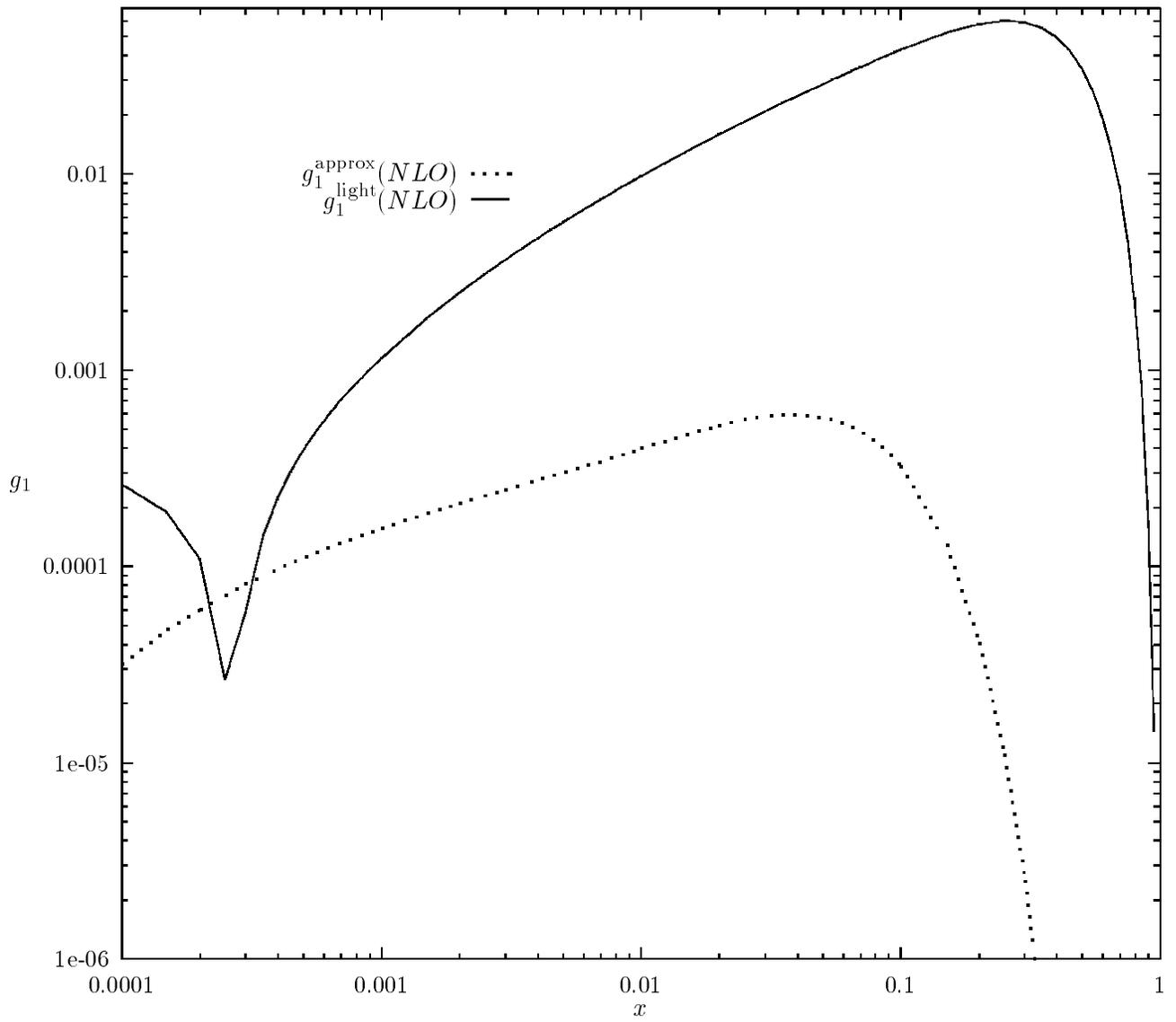
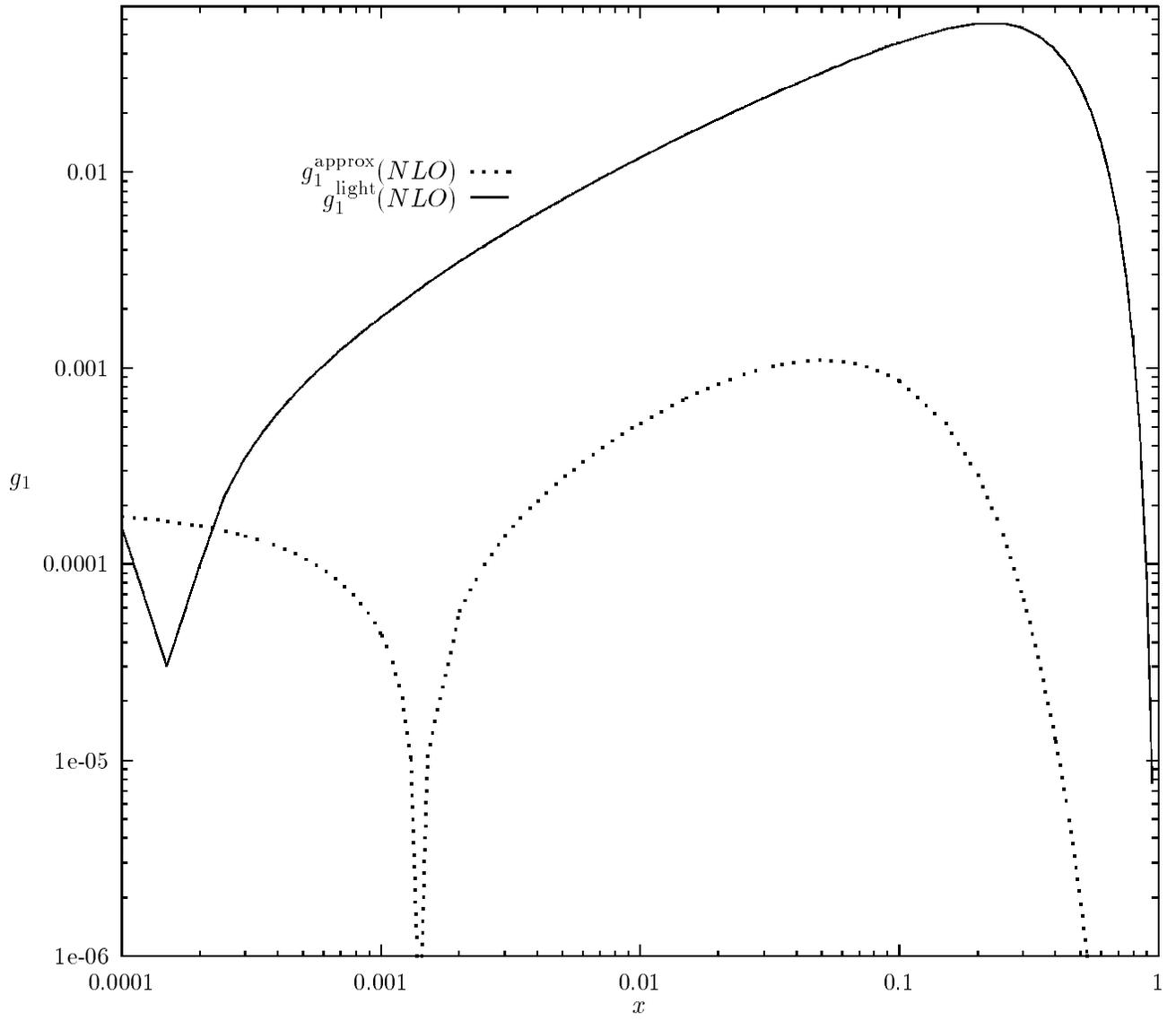
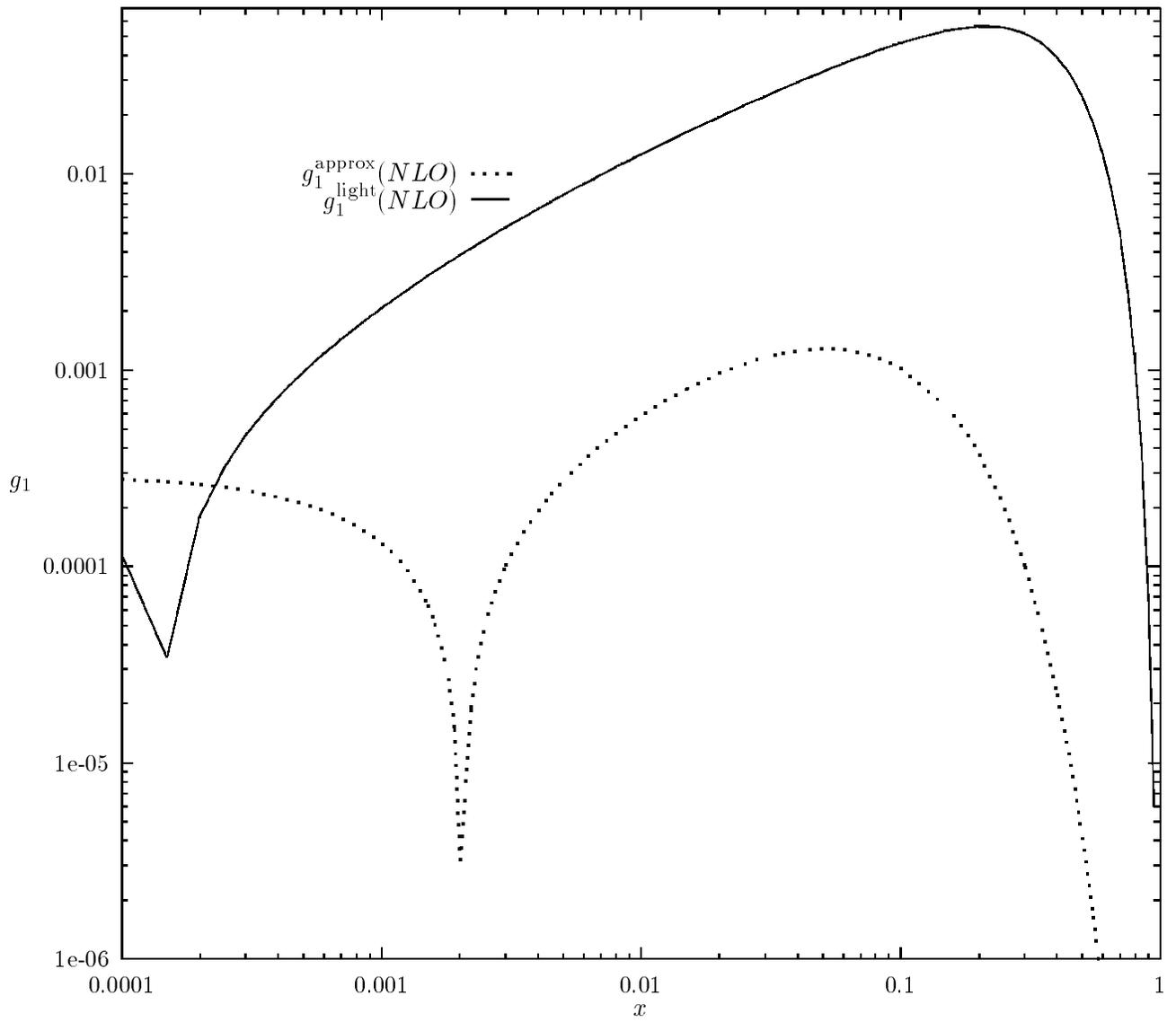
\end{document}